\documentclass[aps,reprint,superscriptaddress,prb]{revtex4-2}
\usepackage{graphicx}
\usepackage{quantikz}
\usepackage{amssymb}
\usepackage{amsmath}
\usepackage{amsthm}  
\usepackage{xurl}
\usepackage{hyperref}
\hypersetup{
    colorlinks=true,
    linkcolor=blue,
    citecolor=blue,
    urlcolor=blue
}
\usepackage{comment}
\usepackage[ruled,vlined]{algorithm2e}
\usepackage[percent]{overpic}
\usepackage{siunitx}
\usepackage{booktabs}
\usepackage[normalem]{ulem}

\newcommand{\figpanel}[2]{Fig.~\ref{#1}(#2)}
\begin{document}
\title{Disorder-enhanced compressibility of Floquet random quantum circuits}

\author{Francesca De Franco}
\affiliation{Forschungszentrum Jülich GmbH, Institute of Quantum Control, Peter Grünberg Institut (PGI-8), 52425 Jülich, Germany}
\affiliation{Institute of Theoretical Physics, University of Regensburg, 93053 Regensburg, Germany}

\author{Dante M. Kennes}
\affiliation{Institut für Theorie der Statistischen Physik, RWTH Aachen, 52074 Aachen, Germany}
\affiliation{Max Planck Institute for the Structure and Dynamics of Matter, Center for Free Electron Laser Science, 22761 Hamburg, Germany}

\author{David J. Luitz}
\affiliation{Physikalisches Institut, Universität Bonn 53115 Bonn, Germany}
\affiliation{Bethe Center for Theoretical Physics, Universität Bonn 53115 Bonn, Germany}

\author{Matteo Rizzi}
\affiliation{Forschungszentrum Jülich GmbH, Institute of Quantum Control, Peter Grünberg Institut (PGI-8), 52425 Jülich, Germany}
\affiliation{Institute for Theoretical Physics, University of Cologne, D-50937 Köln, Germany}

\author{Markus Schmitt}
\affiliation{Forschungszentrum Jülich GmbH, Institute of Quantum Control, Peter Grünberg Institut (PGI-8), 52425 Jülich, Germany}
\affiliation{Institute of Theoretical Physics, University of Regensburg, 93053 Regensburg, Germany}

\date{\today}
\begin{abstract}
Current quantum hardware is limited by noise and decoherence, which restrict the depth of unitary circuits that can be implemented with high fidelity. We investigate how the compressibility of time-evolution operators depends on the dynamical regime of the underlying many-body system. As a testbed, we study a one-dimensional Floquet random circuit with a tunable competition between interactions and on-site disorder. Using tensor-network simulations, we characterize operator growth through the operator-entanglement entropy of the Floquet unitary as well as of out-of-time-ordered correlators (OTOCs). We find rapid operator scrambling at weak disorder, while strong disorder leads to slow OTOC-front propagation and logarithmic or near-logarithmic operator-entanglement growth over the accessible time window. We then optimize shallow brickwall circuits to approximate the Floquet evolution and show that strong-disorder circuits can be compressed to substantially smaller depths than weak-disorder circuits at fixed logarithmic fidelity density. These results suggest that localized or slowly scrambling dynamics provide a favorable regime for compressed quantum simulation on noisy devices.
\end{abstract}

\maketitle
\section{Introduction \label{1}}
Quantum computers offer a universal and controllable framework to model the dynamics of quantum many-body systems. Universality stems from the fact that the sequences of elementary quantum gates would in principle allow for the precise execution of arbitrary unitary dynamics. However, current noisy quantum hardware, i.e. NISQ devices \cite{Preskill2018quantumcomputingin, RevModPhys.94.015004} bear a tradeoff between gate fidelity and circuit depth, rendering long-time simulations hardly accessible. 
As a result, mitigating decoherence effects is a central challenge in realizing practical digital quantum simulations on current hardware. A natural strategy is to approximate a given circuit by a shallower one thereby reducing its susceptibility to noise. This can be achieved by compiling the target evolution into a low-depth brickwall circuit. Such approaches have been successfully applied to the Trotter decomposition \cite{Suzuki1976} of time-evolution operators \cite{PhysRevResearch.6.033062, 10.21468/SciPostPhys.14.4.073}, where higher-order decompositions improve accuracy at the expense of increased gate counts. Recent developments have further advanced variational compression algorithms based on tensor-network (TN) techniques, facilitating the optimization of shallow circuit ansätze acting on hundreds of qubits \cite{Gibbs2025deepcircuit, zhang2024scalablequantumdynamicscompilation, riemannqopt}. 

While general purpose quantum simulation is demanding, some physical models or situations may be particularly suited for investigation on a quantum processor.
In this work, we consider a Floquet random circuit model of many-body localization (MBL)\cite{PhysRev.109.1492, Sierant_2025, PhysRevB.77.064426, PhysRevB.76.052203, PhysRevB.82.174411, 2016JSP...163..998I, PhysRevB.91.081103, Abanin_2017}, which immediately maps to a gate-based computing model.
The Floquet random circuit's MBL-regime is quantified by the spectral statistics of the operator \cite{PhysRevB.105.174205} and by dynamical properties of the system \cite{PhysRevB.98.134204, v4xv-74s7}.
We investigate the compressibility of these circuits, and show that whether a deep circuit admits an efficient low-depth approximation depends on the underlying localization physics of the system. 
Such physical regimes of high circuit compressibility may be particularly suited for optimal digital quantum simulation given finite quantum volumes.

The Floquet random circuit model provides a minimal framework to investigate the competition between interactions and local disorder, allowing us to explore how localization properties impact the compressibility of the circuit. 
MBL arises in disordered Hamiltonian systems, and denotes a regime in which the system fails to thermalize in the thermodynamic limit. Signatures of localization are manifest in the dynamics of the system, which is typically characterized by a slow growth of entanglement entropy and a slow decay of correlations in short-range interacting systems \cite{Sierant_2025}.
To understand how localization affects compressibility of the Floquet random operator, we consider two measures: the time evolution operator complexity, quantified by its operator entanglement entropy (OPEE)\cite{PhysRevB.95.094206, Dubail_2017, PhysRevLett.129.170401} and quantum information scrambling, probed by out-of-time ordered correlators (OTOC)\cite{LarkinOvchinnikov1969, Lieb1972FiniteGroupVelocity, MaldacenaShenkerStanford2016}. Both quantities are sensitive to disorder. Moreover, the minimal structure of the model allows us to easily determine their dependency on the local disorder-tuning parameter within the simulated time windows. 
Studying both quantities, we investigate possible connections between operator complexity growth and information propagation in random Floquet circuits.
We demonstrate that the spreading of OTOCs as an experimentally accessible probe of localization \cite{google2025quantum} is accurately captured by compressed circuits acting on around 100 qubits in the regime of strong disorder.
Therefore, compressed quantum simulation appears as a possible route to address the late time dynamics in supposed regimes of localization.

The manuscript is structured as follows: In Section \ref{2} we introduce the Floquet random model of MBL; In Section \ref{3} we study the circuit dynamics simulated by the time-evolving block decimation (TEBD) algorithm \cite{PhysRevLett.91.147902,SCHOLLWOCK201196}, a state-of-the-art algorithm for one-dimensional systems with short-range interactions. Localization is probed by the OPEE and OTOC observables.
Since rare regions are expected to play a crucial role in the stability of localization~\cite{PhysRevB.105.174205, PhysRevB.108.134204}, we also study sample-to-sample fluctuations of the OTOC front positions. This allows us to identify atypical realizations with faster-than-typical spreading and to probe their connection to local spectral signatures of thermalization. 
Section \ref{4} is dedicated to the variational circuit compression.  
Using a TN-based optimization algorithm similar to that introduced in Ref.\cite{Gibbs2025deepcircuit}, we investigate circuit compressibility as a function of the disorder-tuning parameter. 
Finally, in Section \ref{5}, we summarize our results and discuss the implications of disorder-enhanced circuit compressibility for quantum simulation on near-term quantum hardware.
\section{Circuit model of interaction-disorder competition\label{2}}
The Floquet random circuit as a model of one-dimensional MBL has been introduced in Refs. \cite{PhysRevB.105.174205, PhysRevB.98.134204}. The Floquet operator for one period is given by $U=U_2\cdot U_1$ 
\begin{equation}
\includegraphics[width=.8\columnwidth]{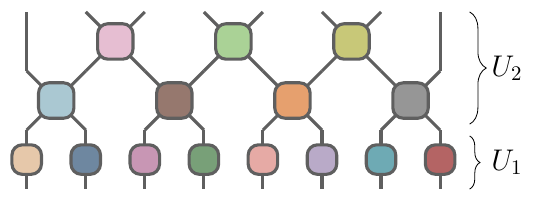}
\label{fig:qc1}
\end{equation}
where $U_1$ is a layer of single-qubit gates and $U_2$ is a brickwork circuit of two-qubit gates.
The set of single-qubit gates $\{d_i\}$ is constructed by independently sampling, for each site $i$, a $2\times 2$ random unitary matrix drawn from the circular unitary ensemble (CUE)
\cite{mezzadri2006generate}. 
We choose the eigenstates of $U_1$ as computational basis. The layer $U_1$,
\begin{equation*}
    U_{1} = d_{1}\otimes d_{2}\ldots\otimes d_{L} 
\end{equation*}
simulates the dynamics induced by local disorder, precisely by random rotations around the Z-axis. 
The brickwall layer $U_{2}$ consists of two-qubit gates $u_{i}$ acting on sites $(i, i+1)$ of the form
\begin{equation*}
    u_{i} = \textnormal{exp}\left( \frac{i}{\alpha}M_{i}\right)  \in \mathbb{C}^{4\times 4}
\end{equation*}
that is sampled from the Gaussian unitary ensemble (GUE), with $M_{i} \in\mathbb{C}^{4\times 4} $ a hermitian matrix; the coupling $\alpha$ $\in \mathbb{R}$ acts as disorder strength. 
At small values of $\alpha$, the dynamics is dominated by the brick-wall layer $U_2$, thus by short-range interactions. Larger values of $\alpha$ reduce the interaction effects, until at very large $\alpha$ (strong disorder regime), the dynamics is dominated by $U_1$. Therefore, $\alpha$ is the disorder-tuning parameter,controlling the interplay between local disorder and interactions.  

At infinite time the averaged spectrum gap ratio indicates a crossover from an ergodic to a finite-size MBL regime at $\alpha_c\approx 6$ for a system of $N=20$ qubits \cite{PhysRevB.105.174205}.
In the following, we investigate how the competition between disorder and interactions affects the transient dynamics of deep Floquet circuits, focusing on operator growth and compressibility.

Throughout the paper, $\langle \ldots \rangle$ denotes an average over independent random-circuit realizations at fixed disorder strength $(\alpha)$.

\section{Quantum information dynamics \label{3}}
Many-body localized states exhibit a slowdown of their dynamics, manifested in both the evolution of correlations and the spreading of quantum information \cite{Sierant_2025}.  
In particular, prototypical MBL systems, such as the random spin-$1/2$ Heisenberg chain predict a logarithmic growth of the state entanglement entropy in the strong-disorder regime \cite{DeChiara_2006, PhysRevB.77.064426, PhysRevB.93.060201, PhysRevLett.109.017202, PhysRevB.90.174302}. 
in sharp contrast to the ballistic entanglement growth observed in chaotic many-body systems \cite{PhysRevLett.111.127205}. 
Similarly, out-of-time ordered correlators displays a qualitatively distinct dynamical behavior: in short-range, chaotic systems, the associated OTOC boundary propagates linearly in time \cite{PhysRevX.8.031058, PhysRevX.8.021013, Lieb1972FiniteGroupVelocity}, whereas in the presence of disorder it spreads sublinearly \cite{PhysRevB.96.020406}, and its growth becomes logarithmic in the strong-disorder limit  \cite{https://doi.org/10.1002/andp.201600332,PhysRevB.95.054201}.

In this work, we characterize the circuit dynamics using two operator-space diagnostics. First, we compute the operator entanglement entropy \cite{PhysRevA.76.032316, Dubail_2017, PhysRevB.95.094206}, a state-independent observable that quantifies operator complexity. We find that its time evolution is analogous to the entanglement entropy of a pure quantum state following a global quench in Hamiltonian systems.
Secondly, we consider the OTOC as detector of quantum information spreading in the circuit. 
Namely, we measure (in operator space) the OTOC between two Pauli $Z$ operators at positions $i$ and $j$: 
\begin{equation}
    C_{ij}(t)= || [Z_{i}(t), Z_{j}(0)] ||_{F}^{2} \ .
    \label{eq:otoc1}
\end{equation}
Where the squared Frobenius norm is defined as $|| A ||_{F}^{2} = \frac{1}{\mathcal{N}}\textrm{Tr}(A^{\dagger}A)$. 

In Subsections \ref{3.2} and \ref{3.3}, we discuss the numerical results of the averaged OPEE and OTOC dynamics. In Subsection \ref{3.4} we analyze sample-to-sample fluctuations of the OTOC front positions, and the association between samples with large front positions at finite-times and thermal inclusions in infinite-size random circuits and long time limit.

In this section the Floquet operator of depth $t$ is represented as a matrix product operator (MPO) \cite{SCHOLLWOCK201196}:
\begin{equation}
    \includegraphics[width=\columnwidth]{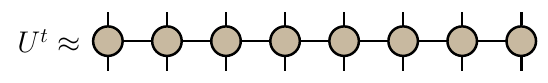}
    \label{fig:MPOapprox}
\end{equation}
with bond dimensions $\chi_{i}$, where $i$ labels the MPO bonds. $U^t$ is approximated using the TEBD algorithm. Details of the MPO construction and truncation scheme are reported in App.~\ref{AppA0}. 
The numerical simulations presented in this manuscript were carried out using code built on the Julia package ITensor \cite{itensor}. 
The source code used to generate the results is available in Ref.~\cite{DeFranco_code}, and the simulation data supporting the figures and analyses are available on Zenodo~\cite{defranco_2026_data}.

\subsection{Operator entanglement entropy\label{3.2}}
We investigate the OPEE as a measure for the complexity of deep Floquet random circuits $(U^t)$.
The OPEE is defined analogously to the entanglement entropy of a pure quantum state: 
an MPO can be regarded as a matrix product state with two copy of physical indices. Hence, the operator entanglement entropy for a cut at position $i$ of an MPO reads \cite{PhysRevB.95.094206, PhysRevA.87.022111}:
\begin{equation}
    \textrm{OPEE}_{i}=-\sum_{k=1}^{\chi_i}\lambda_{k}^{[i] \ 2} \ \textrm{log}_{2}(\lambda_{k}^{[i]\ 2}) 
    \label{opeedef}
\end{equation}
where $\{\lambda^{[i]}\}_{k}$ are the singular values of the MPO at position $i$, with normalization $\sum_{k=1}^{\chi_i}\lambda_{k}^{[i] \ 2}=1$. We analyze the maximum value of the OPEE over all bipartitions of the circuit
\begin{equation}
    \textrm{OPEE}_{\textrm{max}}=\textrm{max}_{i}\left[\textrm{OPEE}_{i} \right]
\end{equation}
We choose the maximum as a summary statistic to account for the inhomogeneous entanglement profiles typically produced by the random circuits. 
When the system reaches thermalization, the maximum OPEE occurs at half the system size, which is consistently observed in circuits with a small number of qubits, $N\approx 10$. 

\subsubsection{Results for 10-qubits circuits}
\begin{figure}[t]
    \centering
    \includegraphics[width=\columnwidth]{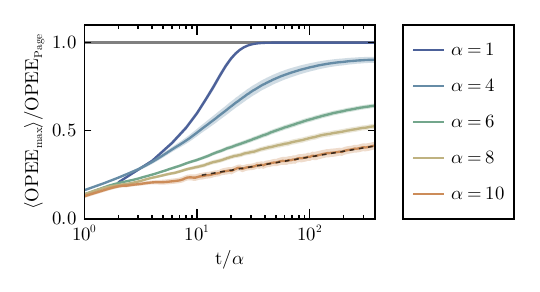}
    \caption{Dynamics of $\langle \textrm{OPEE}_{\textrm{max}}\rangle$ for $N = 10$ qubit circuits in units of the Page value \eqref{pagevalue}. Shaded regions represent the standard error on the mean value. For $\alpha = 1$, the maximum entropy saturates to unity. For $\alpha = 10$, data displays a logarithmic growth over the accessible time window. The number of circuit realizations is reported in Table~\ref{table1}. The dashed line represents the logarithmic growth trend and is included to guide the reader’s eye.}
    \label{fig:opee1}
\end{figure}
We consider a small circuit at first, with $N=10$ qubits. The $\textrm{OPEE}_{\textrm{max}}$ averaged over circuit realizations is shown in Fig.\ref{fig:opee1}. 
The time axis is scaled by the disorder strength $\alpha$, reflecting the observation that for $\alpha > 1$ the early-time dynamics collapse onto a common curve, before interaction-induced deviations emerge.
At $\alpha=1$ the maximum OPEE reaches saturation, with its final value corresponding to the entanglement entropy of a random unitary operator, i.e., the Page value \cite{PhysRevLett.71.1291, PhysRevLett.72.1148}.
Its value for an equal bipartition reads 
\cite{PhysRevB.95.094206}
\begin{equation}
    \textrm{OPEE}_{\textrm{Page}} = N - \frac{1}{2\textrm{ln}2}  + O(4^{-N})
    \label{pagevalue}
\end{equation}
where the entropy is expressed in bits for consistency with the OPEE data in Eq.\eqref{opeedef}.
Saturation is only seen within the simulated time at $\alpha=1$. For $\alpha=4$ and $\alpha=6$ the data are indicative of saturation below the Page value.
Understanding the OPEE dynamics before saturation is particularly relevant: for $\alpha=10$ the data looks consistent with logarithmic behavior. However, at small $\alpha$, the early onset of saturation precludes a quantitative analysis of the dynamics. 

With increasing the number of qubits, $\textrm{OPEE}_{\textrm{max}}$ saturates to its equilibrium value at later times. For hundreds of qubits the saturation time is not accessible numerically, as at small disorder $(\alpha)$ the OPEE growth constrains tensor-network simulations to early-time regimes. 
However, this allow us to determine the characteristic of the growth at small $\alpha$ over time intervals preceding the onset of saturation.

\subsubsection{Results for 100-qubits circuits}
\begin{figure}[t]
    \centering
    \includegraphics[width=\columnwidth]{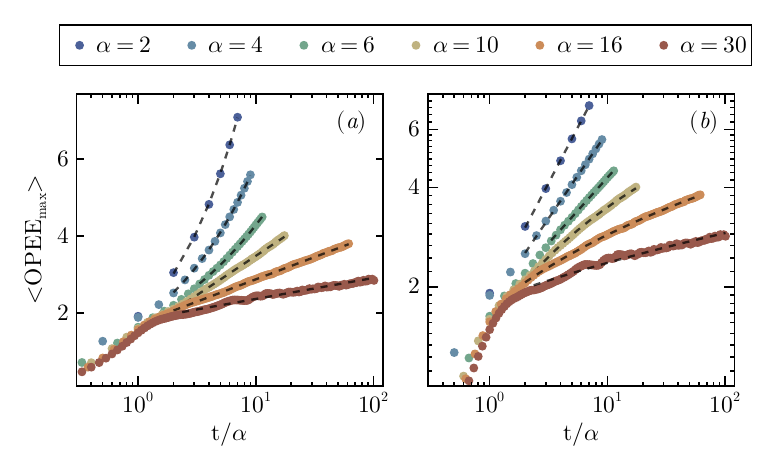}
    \caption{Dynamics of $\langle\textrm{OPEE}_{\textrm{max}}\rangle$ for circuits with $N=100$ qubits. The results are shown in log-linear scale (a) and log-log scale (b). Error bars are within marker size. The dashed lines are fit to the data. $\langle\textrm{OPEE}_{\textrm{max}}\rangle$ is fitted to a power-law $a(t/\alpha)^{\beta}$ for $\alpha \leq 6$, while data for $\alpha\ge 10$ are fitted to $a + b\cdot\textrm{log}(t/\alpha)$. The number of circuit realizations is reported in Table \ref{table2}.}
    \label{fig:opee2}
\end{figure}

Fig. \ref{fig:opee2} shows $\langle\textrm{OPEE}_{\textrm{max}}\rangle$ for circuits with $N=100$ qubits. To better distinguish the dynamics for different disorder strengths, the data are displayed in log-linear scale (Panel a) and log-log scale (Panel b). 
The maximum entanglement entropy shows a rapid increase at small $\alpha$, which limits the accessible time window.  
Despite the high computational cost, the limited time window suffices to identify clearly non-logarithmic growth of the entropy. 

In particular, as shown in ~\figpanel{fig:opee2}{a}, the $\langle\textrm{OPEE}_{\textrm{max}}\rangle$ for $\alpha\leq 6$ displays a pronounced curvature on a log–linear scale, indicating a clear deviation from logarithmic growth. Instead, the data are well described by a power-law behavior $\approx (t/\alpha)^{\beta}$ as shown in ~\figpanel{fig:opee2}{b}, which we fit at the largest times. 
For $\alpha \in [10,30]$, the curves in ~\figpanel{fig:opee2}{b} exhibit weak or nearly vanishing curvature, consistent with an approximately logarithmic dynamics. Given the pronounced dynamical slowdown at strong disorder, it is difficult to distinguish logarithmic from power-law time evolution. Indeed a power-law growth can describe the dynamics as well. In Section \ref{3.3} we estimate the dependence of the power-law exponents on $\alpha$, $\beta=\beta(\alpha)$ and discuss the meaning of the observed behavior.

The $\langle\textrm{OPEE}_{\textrm{max}}\rangle$ dynamics allow us to identify a disorder interval in which the dynamics is clearly non-logarithmic. Specifically, for $\alpha \leq 6$, the averaged $\textrm{OPEE}_{\textrm{max}}$ exhibits an algebraic growth in time, while its dynamics is approximately logarithmic for $\alpha\geq 10$.
These results for Floquet circuits of $N=100$ qubits (within the simulated time) are consistent with the spectrum gap ratio results for circuits of $N=20$ qubits, for which the value of $\alpha$ for the transition from an ergodic to a finite-size MBL regime is $\alpha\approx 6$ \cite{PhysRevB.105.174205}.

\subsubsection{Bond dimension growth\label{3.1}}
\begin{figure}[t]
    \centering
    \includegraphics[width=\columnwidth]{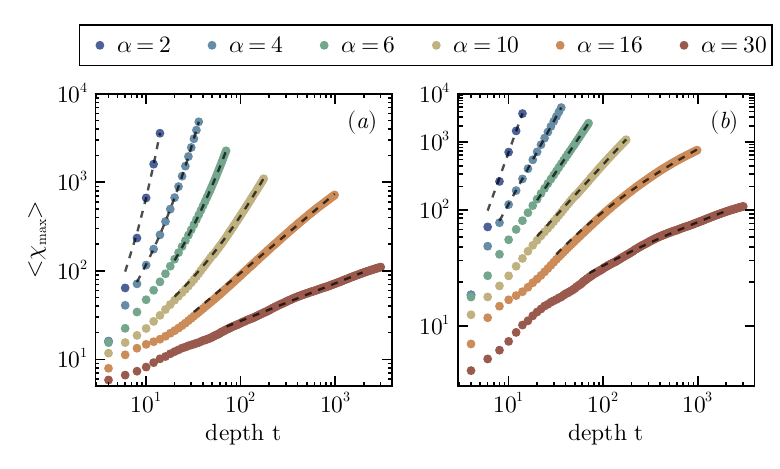}
    \caption{Dynamics of $\langle\chi_{\textrm{max}}\rangle$ for circuits with $N=100$ qubits as a function of circuit depth $(t)$. The number of realizations is reported in Table \ref{table2}, the error bars are within marker size. 
    Panel (a) uses a log-log scale while panel (b) uses double-logarithmic (y-axis) and logarithmic (x axis) scales.
    At $\alpha = 30 $ we can observe a very long time window as the bond dimension is at most $\approx O(10 ^ 2)$. At $\alpha\in [16,30]$ the bond dimension grows as a power-law up to $t=10^3$, consistently with the logarithmic behavior observed for the OPEE for these disorder values. 
    For $\alpha\in [2-10]$ the growth looks like a stretched exponential $\chi\approx a^{t^{b}}$, that is always smaller than $<4^{\lceil t/2\rceil}$ which is the largest possible bond dimension growth (not shown).}
    \label{fig:bond_dimensionN100}
\end{figure}

As a measure of the operator $U^t$ compressibility into an MPO, we display in Fig. \ref{fig:bond_dimensionN100} the maximum bond dimension of the MPO averaged over circuit realizations.
\begin{equation}
\langle\chi_{\textrm{max}}\rangle =\langle \textrm{max}_{i}\chi_{i}\rangle\ .
\end{equation}
For the largest values of $\alpha$, namely $\alpha = 16$ and $\alpha = 30$, the long-time dynamics is compatible with a power-law, clearly visible in the log-log plot of Fig. \figpanel{fig:bond_dimensionN100}{a}. 
For times $t \geq 10^3$, the curve corresponding to $\alpha = 30$ exhibits an additional deceleration, which is why the fit is restricted to the time interval $[70, 2000]$.
Since \(\mathrm{OPEE} < \log(\chi)\), the bond dimension provides an upper bound for the operator entanglement entropy. 
Therefore, the power-law behavior at the highest disorder values of $\langle\chi_{\textrm{max}}\rangle$ is consistent with the slow logarithmic growth that we observe for $\langle\textrm{OPEE}_{\textrm{max}}\rangle$. For the interval $\alpha \in [2,10]$, the time evolution is visually consistent with a stretched-exponential behavior of the form $\chi \approx a^{t^b}$, as highlighted in \figpanel{fig:bond_dimensionN100}{b}, again consistent with the observed growth of the OPEE.

\subsection{Out-of-time ordered correlators\label{3.3}}
Out-of-time-ordered correlators quantify quantum information scrambling, namely the loss of memory of the system’s initial state \cite{Hosur2016,Schnaack2019,Boelter2022}. In short-range interacting systems, this scrambling can travel at most ballistically \cite{MaldacenaShenkerStanford2016,Bohrdt_2017,Mi_2021}.
We consider the OTOC between two Pauli-Z operators, defined as the squared Frobenius norm of their commutator (see the definition in Eq.\eqref{eq:otoc1}).
Expanding the norm and using the unitarity and hermiticity of the $Z$ operator yields
\begin{equation}
    C_{ij}(t) = 1 - \frac{1}{\textrm{dim}(\mathcal{H})}\textrm{Tr}\left( Z_{j}Z_{i}(t)Z_{j}Z_{i}(t) \right) \label{eq:otoc2}
\end{equation}
where $\mathcal{N}=2\cdot\textrm{dim}(\mathcal{H})$. 
The out-of-time correlator vanishes at distances and times for which operators $Z_{j}$ and $Z_{i}\left( t\right)$ are not causally connected, i.e., the two operators commute. We use this fact to improve the efficiency of our numerical calculations as discussed in the following. 

\subsubsection{MPO-based computation of OTOCs}
The operator $Z_{j}Z_{i}(t)Z_{j}Z_{i}(t)$ is a fourth-order term in the time-evolution operator $U^t$, therefore, it requires a large computational cost for classical simulations. 
In practice, each local operator, $Z_i$ and $Z_j$ induces a light cone in the circuit, such that only gates within the overlap of the two light cones are causally connected and therefore contribute to the OTOC.
The operator $Z_jZ_{i}(t)=Z_j(U^t)^\dagger Z_{i} U^t$ is then computed with TEBD, as follows: 
\begin{equation}
    \includegraphics[width=\columnwidth]{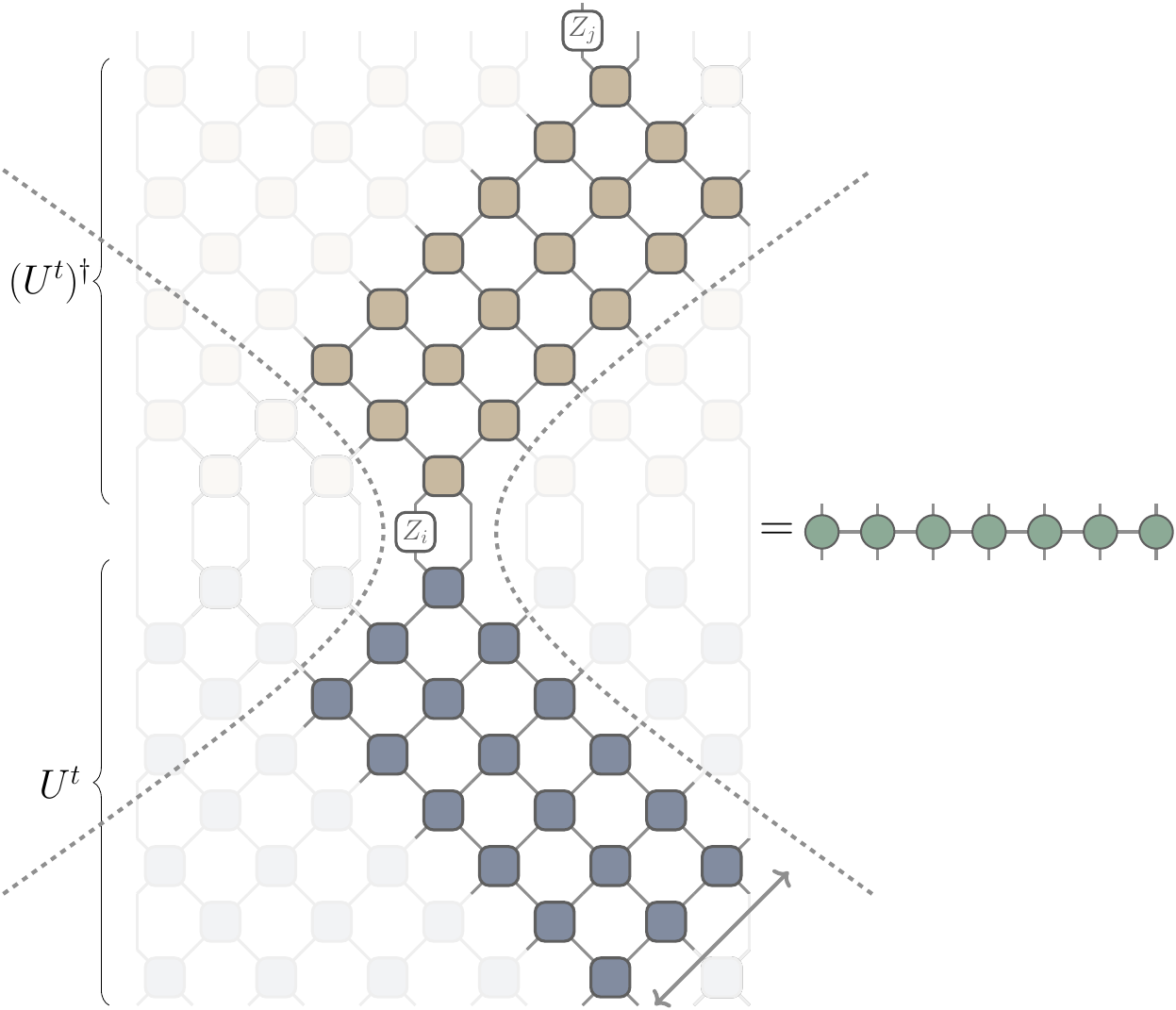}
    \label{fig:otoc0}
\end{equation} 
where only the highlighted gates contribute to the time evolution of the OTOC, since gray gates collapse with their adjoints to unity. In the TEBD simulations these gates are applied as a sequence of staircase circuits.
We denote with $N_{sc}$ the total number of staircase circuits $\in U^t$. 
The resulting MPO parameters scale (worse-case) as $\mathcal{O}(2t\cdot 2^{2N_{sc}})$, where $2t$ is the maximum extent of the MPO spatial support, and $2^{2N_{sc}}$ indicates the bond dimension scaling. We only consider distances $\vert i - j\vert \leq t$, since for larger ones the OTOC is exactly zero. 
We get $N_{sc}= \left\lfloor \dfrac{t - \left| j - i\right| \pm 1}{2} \right\rfloor + 1$ where the sign is negative for $j>i$ and positive for $j<i$. 
(see App.~\ref{AppB}). 
Therefore for $\vert i-j\vert\ll t$ the bond dimension -- in general -- grows exponentially with circuit depth, whereas for $\vert i-j\vert\approx t$ it remains small, making the computation more efficient. 
Beyond the circuit structure, disorder strength also plays a crucial role: at large disorder we expect the upper bound of the bond dimension to be $\chi_{\mathrm{max}}\ll 2^{2N_{sc}}$, enabling the computation of correlators at longer times. 
\subsubsection{OTOCs at weak disorder}
\begin{figure}[t]
    \centering
    \includegraphics[width=\columnwidth]{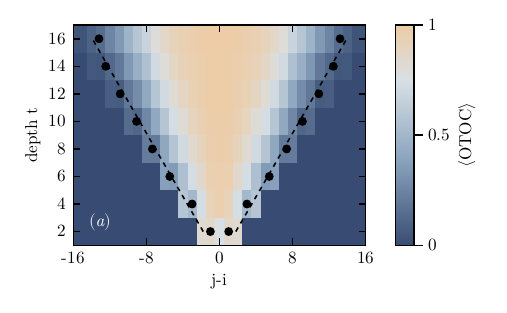}
    \includegraphics[width=\columnwidth]{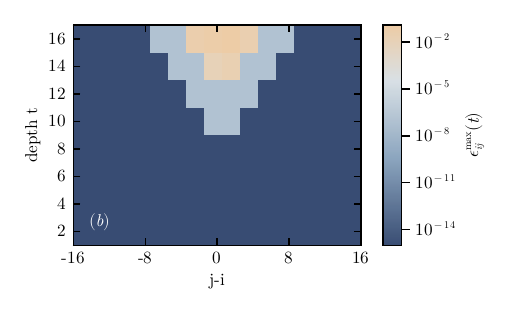}
\caption{(a):$\langle \textrm{OTOC}\rangle $ and relative front position (circles) spreading for $\alpha = 1$. The averaged front error bars are within marker size.
The front position is determined by the OTOC threshold $\eta=0.1$. The power-law ($a\cdot t^\beta$) fit to the front, give an exponent $\beta\approx 1$, and velocity $\approx 0.9$. The number of realizations is reported in Table \ref{table3}. (b):Maximum truncation errors of the MPO $Z_jZ_{i}(t)$ for the data in subplot (a). For each pairs of sites $(i,j)$, $\epsilon^{\max}_{\mathrm{ij}}(t) = \max_{N_c(t)} \left(\epsilon_{ij}(t)\right)$, where $N_c(t)$ denotes the number of circuit realizations at depth $t$. 
Different depth $t$ corresponds to independent simulations. The MPO results from multiple stacking of $N_{sc}$ staircase circuits. For a given $t$, the bond dimension is large at short distances, $\vert i-j\vert\ll t$, and to ensure accuracy, we set the maximum bond dimension at $2^{12}=4096$. With increasing the distance the bond dimension decreases. For given $t$ and distance, plot $(b)$ shows the maximum truncation error over circuit realizations.}
\label{fig:otoc1}
\end{figure}

We first consider circuit realizations at $\alpha=1$. \figpanel{fig:otoc1}{a} shows the average OTOC for circuits of depth up to $t = 16$. 

The data points indicate the front position, that is, the spatial boundary of the correlator. In absence of a universal convention, we define it as the distance at which the squared commutator $C_{ij}(t)$ reaches $C_{ij}(t)=\eta = 0.1$. The front is obtained by averaging the front positions of individual circuit realizations. 

In \figpanel{fig:otoc1}{a}, the left- and right-moving fronts are averaged separately to highlight the linear light-cone. 
In all other analyses, no distinction is made between the two directions. Since the gates are random, the left- and right-moving fronts can be regarded as independent circuit realizations and are therefore combined into a single ensemble.
The front-position data are fitted to a power-law of the form $a\, t^{\beta}$, resulting in an exponent $\beta = 1$ within the associated error bars. The $\beta$-exponent is a priori independent from the power-law exponent of the $\rm{OPEE}$ dynamics at weak disorder. The resulting linear light-cone spreading signals the chaotic dynamics arising in disorder-free random brick-wall circuits.

To assess numerical precision of our simulations, we consider the truncation error, i.e. the norm of discarded singular values of the MPO $Z_{i}(t)$ induced by setting an upper bound to the bond dimension. \figpanel{fig:otoc1}{b} displays the largest truncation error over circuit realizations. 
The highest computational cost comes from computing the OTOC at short distances $\vert i-j\vert\ll t$, where $N_{sc}\propto t$, and thus the bond dimension of $Z_{i}(t)$ grows exponentially in time. The opposite case - the most efficient one - is at $\vert i-j\vert=t$, where $N_{sc}=1$, and $\chi_{Z_{i}}(t)=4$. Therefore, for a given $t$, the truncation error is high at short distances and decreases at larger ones. Hence, the front position computation at disorder $\alpha=1$ is highly accurate, as the OTOC decays to the threshold $\eta$ close to the boundaries $\vert i-j\vert\approx t$.

In the following, we focus on the dynamics of the front position. For intermediate disorder strengths, it is not necessary to compute the OTOC at very short distances, which is the most computationally expensive regime. At strong disorder, however, the effect of disorder on the OTOC boundaries is not known a priori. Nevertheless, in this regime the OPEE grows relatively slowly, following a logarithmic dynamics. We therefore expect reliable accuracy at short distances for bond dimensions smaller than $2^{12}=4096$, i.e., the bond dimension upper-bound used in this subsection. 

\subsubsection{Effects of disorder}
\begin{figure}[t]
    \centering
    \includegraphics[width=\columnwidth]{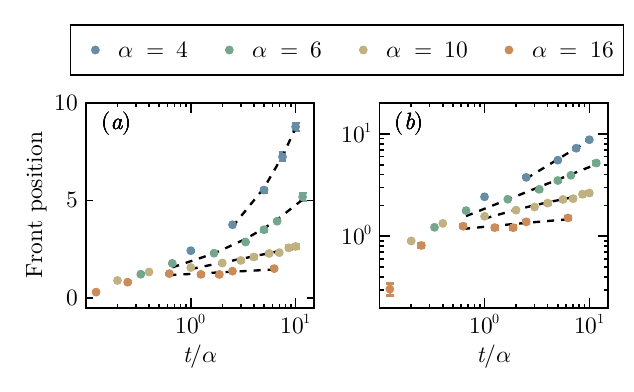}
    \caption{Dynamics of the averaged OTOCs front position for various disorder strengths. The results are shown in log-linear scale (a) and log-log scale (b). The time-axis is in unit of $\alpha$. Error bars indicate the standard error of the averaged values.
    The front position is defined as the distance at which the OTOC has decayed to $90\%$ of its maximum value, that is when $C_{ij}(t) = \eta = 0.1$. The lines represent fits to the numerical data: for $\alpha \leq 6$ we employ power-law fits. For $\alpha \geq 10$, the data are fitted to a logarithmic growth. At strong disorder, deeper circuits are required to determine whether deviations from the logarithmic growth eventually emerge. The number of realizations is reported in Table \ref{table3}.
    }
    \label{fig:otoc3}
\end{figure}
 Fig. \ref{fig:otoc3} displays the dynamics of the front position averaged over circuit realizations for discrete disorder
 values varying from $\alpha=4$ up to $\alpha=16$.    
The average OTOC front position at $\alpha>1$ exhibits a spatial spreading ranging from an algebraic behavior observed at $\alpha=4$ and $\alpha=6$, with a time dependence compatible with a sublinear growth $\sim (t/\alpha)^{\beta(\alpha)}$; to a logarithmic growth observed at $\alpha \geq 10$ (the fits to the data are shown as dashed lines in Fig. \ref{fig:otoc3}). 
\figpanel{fig:otoc3}{b} shows the data in log-log scale, highlighting the algebraic growth in time for $\alpha\leq 6$. 

$\langle\textrm{OPEE}_{\textrm{max}}\rangle$ and the averaged OTOCs front time evolution exhibits compatible behavior as functions of $\alpha$, both being described by an algebraic growth at intermediate disorder and an approximately logarithmic one for $\alpha\geq 10$. 
\subsubsection{OPEE-OTOC association}
\begin{figure}[t]
    \centering
    \includegraphics[width=\columnwidth]{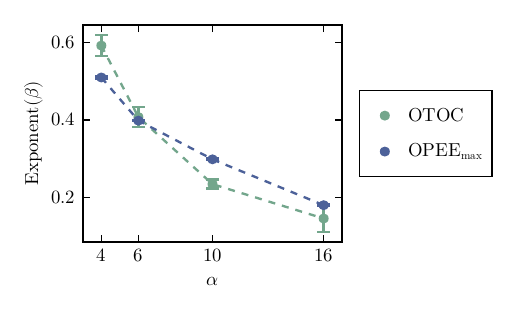}
    \caption{Exponents $\beta$ extracted from the power-law fits of the averaged $\textrm{OPEE}_{\textrm{max}}$, and the averaged OTOC front positions with threshold $\eta=0.1$. The exponents are fitted over comparable time intervals, the details are reported in Table \ref{table4}. These quantities display a similar algebraic time evolution.}
    \label{fig:otoc4}
\end{figure}

In Fig. \ref{fig:opee2} and Fig. \ref{fig:otoc3} we show that at strong disorder both OPEE and OTOC dynamics are consistent with logarithmic growth. 
However, within the accessible time window the dynamics is sufficiently slow that a weak algebraic growth yields a comparable description. Resolving the asymptotic form would require substantially longer times, where deviations from logarithmic propagation could become manifest, but this regime is hardly accessible with TEBD simulations.

Since an algebraic form provides a good description of the data, we use it as a common finite-time parametrization across disorder regimes. Figure~\ref{fig:otoc4} shows the resulting exponent $\beta(\alpha)$, extracted from power-law fits $\sim t^\beta$ of the averaged maximum OPEE and OTOCs front position, for disorder strengths $\alpha>1$.
Notably, the two sets of data are compatible for each value of $\alpha$, suggesting that the two observables are strongly associated for intermediate to strong disorder values.
Furthermore the slow decay of $\beta(\alpha)$ indicates the crossover between sublinear and logarithmic growth appears smooth rather than sharp. 

The OTOC front provides a natural estimate of the spatial region over which initially local operators have developed significant support, and therefore gives an argument for the similar $\alpha$-dependence of OPEE and front growth. 
This finding is particularly relevant because the OTOC can be measured experimentally and may therefore serve as a proxy for operator complexity as quantified by OPEE.

\subsection{Sample-to-sample fluctuations: are there thermal inclusions? \label{3.4}}
The results discussed so far characterize the dynamics at the level of averaged quantities. However, averaged observables may wash out rare-event fluctuations across circuit realizations.
We analyze sample-to-sample fluctuations of the OTOC-front measures. This allows us to assess how broadly distributed the front positions are and whether rare disorder configurations behave atypically. 
\begin{figure}[t]
    \centering
    \includegraphics[width=\columnwidth]{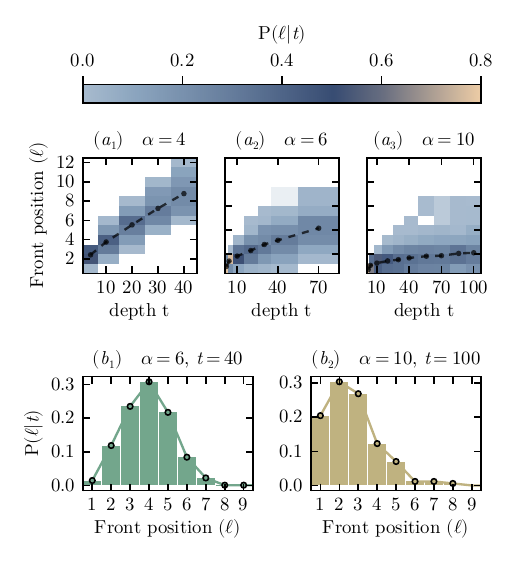}
    \caption{Top panel: Heatmaps for $\alpha\in[4, 6, 10]$ of the front-position distribution $P(\ell|t)$, obtained from circuit realizations independently sampled at each depth $t$; the sample counts are reported in Table~\ref{table3}. Each column is normalized separately, so that $\sum_{\ell} P(\ell|t)=1$ for every $t$. The black lines indicate the averaged values. Lower panel: Front-position distribution for $(\alpha=6, t=40)$ on the left and $(\alpha=10, t=100)$ on the right.}
    \label{fig:otoc56}
\end{figure}
\begin{figure}[t]
    \centering
    \includegraphics[width=\columnwidth]{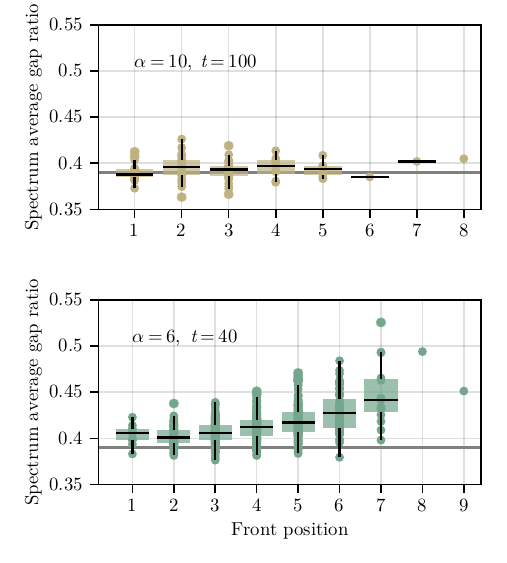}
    \caption{Spectrum average gap ratio (y-axis) of sub-circuits acting on $n=10$ qubits against the OTOC-front (x-axis) measured at a given depth $t$. The boxplots indicate the gap ratio distribution ($\bar{r}_\ell$) for a subset of samples $\mathcal{S}_{\ell}$ having the same front position ($\ell$). The circle dots represents the values ($\bar{r}_\ell, \ell$) for the individual samples.
    }
    \label{fig:otoc7}
\end{figure}
\figpanel{fig:otoc56}{b} shows the probability distribution of the front position for $(\alpha=6,t=40)$ and $(\alpha=10,t=100)$. In both cases, and especially at strong disorder $\alpha=10$, the tail of the distribution is characterized by a small fraction of samples whose front position is substantially larger than the average.
The top panel of Fig. \ref{fig:otoc56} shows the front position statistics over time for three disorder values; the number of realizations is reported in Table \ref{table3}. For $\alpha=4$, the samples spread at the same rate of the average, $\approx t^{0.6}$. For both $\alpha=6$ and $\alpha=10$, we observe pronounced sample-to-sample fluctuations. 
Individual realizations may display a plateau at late times which prevents the extraction of their functional form; e.g., this is the case for the maximum front-position spreading in Panel $(a_3)$ -- which corresponds to the front dynamics of a single circuit realization --, it initially grows rapidly before reaching a plateau. In view of the intrinsic difficulty associated with characterizing sample-to-sample fluctuations at finite times, we adopt an alternative strategy: we employ an infinite-time diagnostic. 

Understanding the fate of the outliers in the infinite-time limit is of interest, as these samples could be candidates for thermal inclusions, i.e. sub-regions of the circuit that in the thermodynamic limit act as thermal baths for the whole system \cite{PhysRevB.95.155129}.
We emphasize that in our analysis the system size does not play a role. Indeed, the front position is always smaller than the circuit depth $t$; furthermore, because of the stochastic nature of the gates, each OTOC realization can be interpreted as an OTOC with light-cone centered on a qubit of an effectively infinite chain. This allows us to consider individual circuit realizations as sub-regions of an infinite-size circuit. The volume of a sub-region is determined by its maximum front position expansion, as this volume encompass the number of qubits that are causally connected. 

Within this framework, we classify individual samples as thermal or localized by comparing the front position at finite time with an infinite-time probe of thermalization, that is the spectrum gap ratio. 
We consider the front position distributions at $(\alpha=6, t=40)$ and at $(\alpha=10, t=100)$ (\figpanel{fig:otoc56}{b}). For these ensembles the maximum front position expansion occurs over distances of fewer than \(n = 10\) qubits. Therefore, for each sample we consider the sub-region of $n=10$ qubits containing the OTOC light-cones, and compute its spectrum-averaged gap ratio.
We indicate with $\{ \bar{r}_{\ell}\}$ the spectrum-averaged gap ratio for a subset $\mathcal{S}_{\ell}$ of circuit realizations with front position $\ell$.

At $\alpha=10$, there is no correlation between the average gap ratio and the front position: independently of the growing front, the gap ratio values are close to the Poisson value (horizontal line) of $\sim 0.39$, indicating that the sub-regions considered are localized. 
The behavior is different for $\alpha=6$, where the gap ratio $(\bar{r}_{\ell})$ varies significantly from sample to sample, and its maximum increases with $\ell$. The maximum value it can acquire is $\bar{r}_{\ell}\sim 0.59$, as predicted from random matrix theory. 
For $\ell \leq 7$—where the statistics remain reliable, while larger values of $\ell$ are affected by too few occurrences—realizations with the same finite-time front position can nevertheless display different spectral properties. 
This observation suggests that the distinction of thermal and non-thermal samples is revealed in the spectral properties of the subregion enclosed by the OTOC light-cone. 
This strategy for the classification of thermal and non-thermal samples is of interest in the regime of intermediate disorder strength ($\alpha\approx 6$ in our model), where the spectrum statistics, strongly dependent on the system size, may indicate an MBL regime for the full systems, while dynamical properties are not MBL-like. 
\subsection{Implications for compressibility}
The results of this section have a direct implication for circuit compression. So far, we used a MPO compression of the Floquet circuit $U^t$ put forward with TEBD simulations for gate evolution. This MPO-based approach allowed us to monitor the growth of operator entanglement and to compute OTOC fronts, thereby quantifying both operator complexity and information spreading. 
Their slowdown being correlated with increasing disorder suggests that, in the strong-disorder regime, the Floquet unitary should admit a shallower circuit representation than in the weak-disorder regime. 

It is useful to stress, however, that the OPEE and OTOC calculations are limited by different computational bottlenecks. The OPEE can be extracted directly from the singular-value structure of the MPO representation of $U^t$. The OTOC requires the contraction of local operators within overlapping light cones. A priori, all causally connected two-qubit gates inside this spacetime region contribute to the correlator, which restricts the accessible depths and distances in the direct MPO--TEBD calculation.

In the next section, we turn to direct circuit compression as an alternative approach to render long times accessible with limited computational resources.

\section{Variational circuit compression \label{4}}
Motivated by the connection between slow operator growth and compressibility discussed above, we now test directly whether the Floquet evolution $U^t$ can be approximated by a shallower circuit. 
Instead of working with the original depth-$t$ Floquet circuit, we variationally approximate $U^t$ by a low-depth brick-wall circuit with the same local connectivity but fewer layers.
Such shallow circuits are naturally tailored for digital simulations on NISQ devices, whose relevant limitation is the circuit depth, since deeper circuits are more strongly affected by decoherence and gate errors. Therefore, the study of long-time dynamics in disordered and potentially MBL systems admitting compressed circuit simulation may be a natural target for near-term quantum hardware.

For the compression problem, we consider Floquet circuits of $N=100$ qubits, matching the system size used for the MPO-compression in the previous section. The variational problem consists in optimizing a suitable cost function, introduced in the next section. Next, we discuss the optimization algorithm. 

As we outline in the following, the variational compression is based on formulating a suited cost function and minimizing it to find an optimal approximation of $U^t$ with the given circuit depth.
We found that the gate-local gradient-free optimization approach introduced in Ref.~\cite{Gibbs2025deepcircuit} outperforms gradient-based minimization of the cost function as described in Ref.~\cite{Kotil_2024, riemannqopt}. 
Therefore, our focus is on the local optimization approach and we defer a presentation of the gradient-based method as well as a comparison between the two approaches to App.~\ref{AppE}.

\subsection{Cost function} 
A standard choice for the cost function is the fidelity between the target operator $(U^t)$ and an Ansatz circuit $(C)$ of reduced depth,
\begin{equation}\label{fidelity}
F = \dfrac{1}{2^{N}}\left|\text{Tr}\left[\left(U^{t}\right)^{\dagger}C\right]\right|.
\end{equation}
This quantity equals unity when the two operators coincide up to a global phase. However, as in the case of state fidelity \cite{Zhou_2008, math10060940}, $F$ exhibits exponential scaling with the system size $N$, such that small local discrepancies result in an exponentially suppressed global fidelity (The scaling argument is outlined in App.~\ref{AppC}.)

 To remove the scaling with $N$, we use
 as cost function the logarithmic fidelity density
\begin{equation}
    f
    =
    -\frac{1}{N}\log_{2} F
    \label{logf}
\end{equation}
which is expected to be independent of the system size. The variational problem is then formulated as the minimization of $f$. 
\subsection{Optimal compression algorithm}
The Ansatz circuit $C$ consists of random initialized two-qubits gates $G^{t}_{i,i+1}$ arranged in a brickwork structure. $t$ denotes the circuit-layer, and $(i,i+1)$ specify the pair of neighboring qubits on which the gate acts.
To compute the full contraction:
\begin{equation}
    \includegraphics[width=\columnwidth]{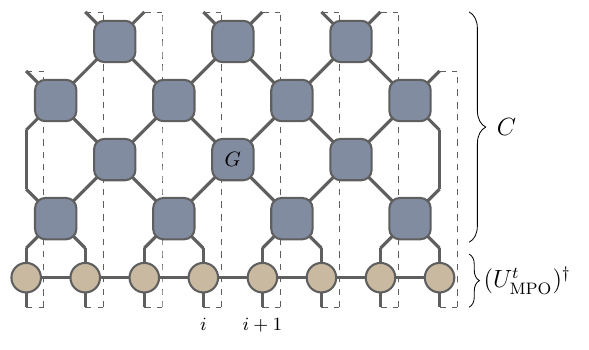}
    \label{fullcontrnet}
\end{equation}
the Floquet operator $U^t$ is represented as an MPO, with its bond dimension chosen so that the logarithmic fidelity density with respect to the exact operator satisfies $f_{\rm{init}}(U^t,\  U^{t}_{\mathrm{MPO}})\leq 10^{-6}$. 
This ensures an accurate initial representation while moderately reducing the computational cost and memory requirements. 
The optimization tsubsequently minimizes $f(U^{t}_{\mathrm{MPO}}, C)$, for which we use a stopping threshold larger than $10^{-6}$, typically of order $10^{-3}$. 
Since $f_{\rm{init}}\ll f$, the error introduced by the MPO initialization is negligible compared to the optimization tolerance.
\begin{algorithm}[ht]
\caption{Floquet circuit compression}
\label{alg:floquet_compression}

\KwIn{Initial target MPO $U_{\rm MPO}^{t}$, initial Ansatz circuit $C$,
threshold $\epsilon$, maximum target depth $t_{\max}$}
\KwOut{Optimized compressed circuit $C$}

\While{$t \leq t_{\max}$}{
    Optimize the Ansatz $C$ with respect to $U_{\rm MPO}^{t}$\;
    $C \leftarrow \mathrm{Optimize}(U_{\rm MPO}^{t},C)$\;

    \eIf{$f \leq \epsilon$}{
        Increase target depth to $2t$ via TEBD\;
        $t \leftarrow 2t$\;
        $U_{\rm MPO}^{2t} \leftarrow \mathrm{TEBD}(U^{t},U_{\rm MPO}^{t})$\;
    }{
        $l\leftarrow$ new layer with random two-qubit gates\;
        Append $l$ to the Ansatz circuit\;
        $C \leftarrow C\cdot l$\;
    }
}

\Return{$C$}\;
\end{algorithm}

The compression routine is illustrated in Alg.\ref{alg:floquet_compression}, where the target evolution depth and the Ansatz depth are increased adaptively according to the compression error.
\subsubsection{Local updates}
The optimization of $f$ is carried out via local updates of the Ansatz $C$. A single optimization step consists of sequentially updating the two-qubit gates $G^t_{i,i+1}$, following a prescribed update order through the circuit. 
The scheme involves an outer sweep over the circuit depth (time direction) and an inner sweep over the gates within a layer at depth $t$.

To update $G^t_{i,i+1}$ we minimize the cost function $f$ (eq.\eqref{logf}) with respect to that gate. 
This minimization is equivalent to maximizing the full contraction
\begin{equation}
    \left|\text{Tr}\left[(U^{t}_{\mathrm{MPO}})^{\dagger}\cdot C\right]\right\vert = \left\vert\text{Tr}\left[E\cdot G\right]\right\vert 
\end{equation}
which in particular we express explicitly in terms of the target gate $G$ (here we omit positional indices for clarity).
The \emph{environment} $E$ is a rank-4 tensor obtained by removing $G$ (leaving a hole) from the network in \eqref{fullcontrnet} and contracting all remaining tensors.
The solution to this problem is known exactly, and it is given by the polar decomposition of the environment. 
Specifically, the update $G$ comes from the SVD of the adjoint environment $E^{\dagger}$ \cite{PhysRevB.79.144108}, as follows: 
\begin{equation}\label{localupdates}
E^{\dagger} = WSV^{\dagger}, \qquad G\leftarrow WV^{\dagger}    
\end{equation}
The contraction of the environment $E^{t}_{i, i+1}$ is the most expensive part of the algorithm. For this task we adopt a tensor network-based scheme similar to the one introduced in Ref.\cite{Gibbs2025deepcircuit}. The method consists of encoding the networks above and below the hole as MPOs, so that the associated parameters can be controlled during optimization. Further details are provided in App.~\ref{AppD}.
\subsubsection{Target approximation \label{4.1.2}}
\begin{figure}[t]
    \centering
    \includegraphics[width=\columnwidth]{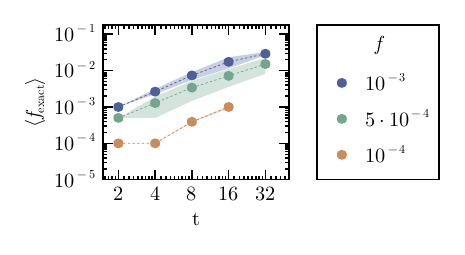}
    \caption{Log-fidelity density \(\langle f_{\mathrm{exact}}\rangle \) between the exact target MPO (eq.\ref{exactarget}) and the optimal circuit obtained from the compression of the approximated target (eq.\ref{targapprox}). We show three circuit realizations, for \(N=100\) qubits and disorder strength \(\alpha=10\). The markers denote the mean, shaded regions indicate the interval between the corresponding minimum and maximum values.
    Distinct colors denotes different stopping threshold $f$ used in the optimization of the approximated target \(U_{\mathrm{MPO}}^{2t} \approx C \cdot C\), with \(C\) the optimal compression of \(U^t\). 
    Although smaller stopping thresholds reduce the overall error, the growth rate of the accumulated approximation error remains similar.
    }
    \label{fig:fidcheck}
\end{figure}
The manipulation of the MPO representation of $U^t$ becomes increasingly costly as the depth grows. To reach longer times, we can take advantage of the fact that the dynamics is generated by repeated applications of the same Floquet unitary. 
In particular, once the circuit at time $t$ has been compressed, yielding an optimized approximation $C_t$ to $U^t$, the evolution over twice the time can be obtained by composing the evolution over time $t$ with itself,
\begin{equation}
    U^{2t}=U^t U^t .
    \label{exactarget}
\end{equation}
Rather than constructing and compressing the exact operator $U^{2t}$, we therefore use the already compressed circuit as a building block and approximate the target at the next time scale as
\begin{equation}
    U^{2t}\approx C_t C_t .
    \label{targapprox}
\end{equation}
This provides a recursive compression strategy, where the compressed circuit at time $t$ is used to generate an effective target for the compression at time $2t$.
Using a target approximation as the depth grows substantially reduces the computational cost by keeping the circuit representation of the target compact. However, the approximation introduces a truncation error that accumulates with each recursion step. 

To quantify this effect, we introduce $f_{\mathrm{exact}}$, which denotes the logarithmic fidelity density between the optimized circuit, obtained from the approximated target constructed according to Eq.\eqref{targapprox}, and the MPO representation of Eq.\eqref{exactarget}.
Figure \ref{fig:fidcheck} shows $\langle f_{\mathrm{exact}}\rangle$, averaged over three circuit realizations, for several optimization thresholds $f$. As expected $\langle f_{\rm{exact}}\rangle = f$ at $t=2$, before any recursive approximation is introduced. At longer times, the discrepancy grows systematically due to error accumulation. 
Reducing the threshold from $10^{-3}$ to $10^{-4}$ lowers the overall error but does not qualitatively change its growth with depth.
The resulting errors remain controlled, although they accumulate with time, leading to an increase of about one order of magnitude over the time window $t\in[2,32]$.
These results support the hybrid strategy used below: we optimize against the exact MPO target at short times and switch to the recursive approximation at late times, where it enables access to longer evolutions while maintaining a compact circuit representation. The accuracy of the optimized circuits is then assessed both through the cost function and through physical observables.

\subsection{Compression results for 100-qubits 
circuits}
\begin{figure}[t]
    \centering
    \includegraphics[width=\columnwidth]{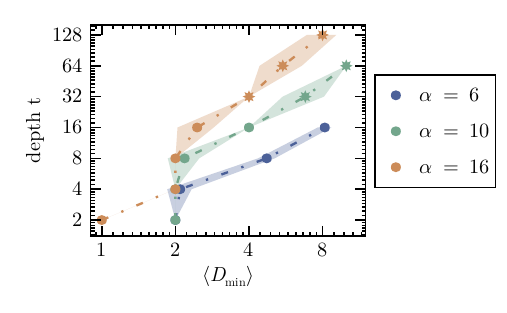}
    \caption{Exact circuit depth $(t)$ against the minimal depth $\langle D_{\mathrm{min}}\rangle$ averaged over circuit realizations, achieved at stopping threshold $f=5\cdot 10^{-3}$. Data are averaged over a moderate ensemble of 11 circuits. The optimizations are performed with a maximum of $3000$ sweeps. The error bars represent the $97.5\%$ confidence interval of the mean, computed using the Student's t-distribution to account for the small sample size. For $t \geq 32$, the target circuit is approximated using the recursive procedure described in the previous section.}
    \label{fig:vcc-res1}
\end{figure}
\begin{figure*}[t]
    \centering
    \includegraphics[width=\linewidth]{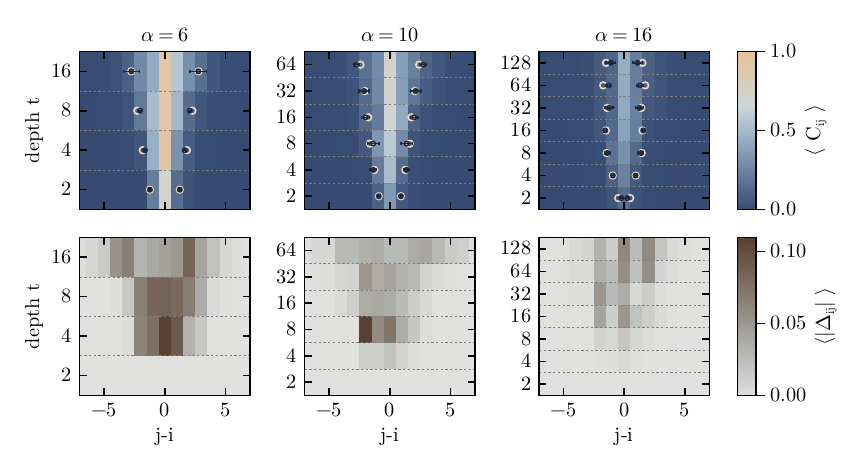}
    \caption{Out-of-time ordered correlators accuracy check for compression results at $\alpha=6$ (first column), $\alpha=10$ (second column) and $\alpha=16$ (third column). Top panel: Plot of the exact OTOC at distance $j-i$ and depth $t$. The points indicate the averaged front position for the exact OTOC, (white dots -- error bars are within marker size --) and for the OTOC of the compressed circuit (blue dots). Bottom panel: The absolute error is between the exact OTOC and the one of the optimal circuit.}
    \label{fig:vcc-res3}
\end{figure*}

Fig.~\ref{fig:vcc-res1} displays the average minimum depth $\langle D_{\rm{min}} \rangle$ achieved through Floquet circuit compression. We compress Floquet circuits of fixed depth $t$ and then average the minimum depth $D_{\rm{min}}$ of the resulting optimal circuits.
We retain only optimization for a logarithmic fidelity density threshold $f=5\times10^{-3}$ (For completeness, in App.~\ref{AppE1} we also show the compression results for the lower threshold $f=10^{-3}$).

We emphasize that $f$ is a local fidelity density. The accuracy of the simulation is assessed through physical observables rather than the fidelity alone. As shown in Fig.\ref{fig:vcc-res3}, the chosen compression threshold is sufficient to accurately capture the quantities of interest.
For sufficiently large depths, the MPO representation of $U(t)$ becomes too costly to use directly within the optimization scheme. Already at $t=32$, its average bond dimension reaches $\chi = \mathcal{O}(500)$, making the computation of the local environments (see App.~\ref{AppD}) particularly expensive as it scales with $\mathcal{O}(\chi\ 4^{\lceil D/2\rceil})$, where $D$ denotes the depth of the Ansatz circuit. Therefore, at late times we exploit circuit periodicity and use the target approximation introduced in the previous section.  
For the largest $\alpha$ considered, $\alpha=16$ (deep MBL), we are able to successfully compress circuits of size $N=100$ of effective depth $\approx\mathcal{O}(10^2)$. 

Compressibility at strong disorder is enhanced by the underlying non-chaotic dynamics. As $\alpha$ is reduced, fitting large and deep target circuits becomes substantially harder, consistent with the faster growth of operator complexity. For the value $\alpha=6$ the results are displayed up to depth $t=16$, where the minimal depth achieved for the optimal circuit is reduced only by a factor of two with respect to the target depth. 

Eventually, the minimal depth quantum circuit bears two truncation errors, one is the minimum threshold $f$, the second is the error introduced by the target approximation when applied. 
To assess the reliability of our results, we compute the averaged OTOC for the optimized circuits and compare it to that of the exact target circuits. The corresponding results are shown in Fig.\ref{fig:vcc-res3}. Each column corresponds to a distinct disorder value. The first row shows the averaged OTOC of the exact circuit, whereas the second row displays the averaged absolute OTOC error. With
\begin{equation}
    \Delta_{ij} = C_{ij}(t) - \tilde{C}_{ij}(\rm{D_{min}})
\end{equation}
denoting the difference between the exact and approximated OTOC.
The absolute error ranges from zero up to $\approx10^{-1}$. At the front position—the distance of interest—the OTOC error is approximately $5\%$. Particularly at $\alpha=16$, the error is stable around $\approx 5\%$, in the full time interval. 
Meaning that shallow circuits of depth $\langle\rm{D_{min}}\rangle\simeq 7$ accurately reproduce the OTOC space-time expansion of deep circuit with $t=128$ layers.
This results enhance OTOC computation in the strong disorder regime, by reducing the gates count in the MPO-based OTOC scheme (view Eq. \eqref{fig:otoc0}) by a factor of $\simeq 18$.

Finally, for $\alpha\lesssim6$, the algorithm struggles to find a good low-depth circuit approximation. This is due to the larger complexity of the operator. In Appendix ~\ref{AppF} we show the complexity of one period of a single circuit-realization of the Floquet circuit $(U)$ trough t-SNE \cite{JMLR:v9:vandermaaten08a}. This method maps the circuit into a two-dimensional space allowing a direct visualization of circuit complexity. Indeed, we observe that, at $\alpha > 1$, the support of the circuit in a two-dimensional space is particularly sensitive to the role of the layer of one-qubit gates $(U_{1})$ simulating local disorder, and to the effect of the disorder strength $\alpha$.  

\section{Discussion and Outlook \label{5}}
We have investigated Floquet random circuits as a platform for compressed quantum simulation, focusing on dynamical signatures of localization. Using the operator entanglement entropy (OPEE) and the out-of-time-ordered correlator (OTOC), we characterized respectively the growth of operator complexity and the spreading of quantum information. These probes exhibit consistent behavior: in the presence of strong disorder, both quantities display slow growth, reflecting constrained dynamics and limited operator spreading.

The suppression of operator complexity, in turn, enables efficient compression of the dynamics, extending the reach of classical simulation. At the same time, the OTOC provides a directly accessible observable within such compressed representations. However, the slow propagation of the OTOC front implies that increasingly large spacetime regions must be retained at long times, leading to a growing computational cost. This interplay identifies a regime in which classical simulation becomes challenging, suggesting a potential window for quantum advantage.

An important direction is the extension to higher-dimensional circuits. Recent results \cite{google2025quantum} have demonstrated quantum advantage in computing OTOCs in two-dimensional random circuits, where tensor network methods are significantly less effective. Extending our approach to such systems would provide a natural setting to further explore this regime.

\begin{acknowledgments}
This work was supported by the Deutsche Forschungsgemeinschaft (DFG, German 
Research Foundation) under Germany’s Excellence Strategy Cluster of Excellence 
Matter and Light for Quantum Computing ML4Q (EXC 2004, project-id 390534769), by 
CRC1639 NuMeriQS (project-id 511713970) and CRC TR185 OSCAR (project-id 277625399).

The authors gratefully acknowledge computing time on the supercomputer JURECA\cite{JURECA} at Forschungszentrum Jülich.
\end{acknowledgments}

\appendix
\section{MPO/TEBD methodology and numerical implementation \label{AppA0}}
In Section \ref{3} the time-evolved Floquet operator is represented as an MPO constructed using the TEBD algorithm. In this appendix, we provide additional implementation details. 
The MPO representation of the full-rank operator is a decomposition into a product of rank-4 tensors (see Eq. \ref{fig:MPOapprox}). The number of MPO parameters scales as $\mathcal{O}(N\chi^{2})$, where $\chi$ denotes the bond dimension. Since the Floquet random circuit is not translationally invariant, the bond dimension may vary from bond to bond. 
The MPO has coefficients in the chosen basis having MPS form. Consequently, the canonical-form construction for MPS tensors extends directly to MPOs \cite{SCHOLLWOCK201196}.
$U^t$ is initialized with an identity MPO ($\chi = 1$), which is evolved with TEBD by sequentially applying the two-qubit gates of one Floquet period $U=U_2\cdot U_1$. For the gate evolution, we first absorb the layer $U_1$ of one-qubit gates in the brickwall structure, hence, we modify the two-qubits gates in the first layer of $U_2$ as $u_{i}\xleftarrow{}u_{i}\cdot \left[ d_{i} \otimes  d_{i+1}\right]$. During gate evolution, the MPO is maintained in mixed canonical form, with the orthogonality center placed at the bond where the currently applied two-qubit gate is absorbed.

To control the bond dimension growth, we use a singular-value decomposition (SVD) truncation cutoff of $10^{-7}$, corresponding to a maximum relative discarded weight of $10^{-7}$ for each bond. The reported results of the averaged OPEE in Subsection \ref{3.2} are not constrained by any maximum bond dimension. Furthermore, reducing the truncation threshold below $10^{-7}$ does not produce any visible changes in the plotted data. The truncation error affects the unitarity of the MPO which is preserved only approximately with an accuracy controlled by the SVD truncation threshold. 

For the OTOC evaluation of Eq.~\eqref{eq:otoc2} in Section~\ref{3.3}, the TEBD evolution of $Z_j(t)Z_i$ is performed using an SVD cutoff of $10^{-7}$ and by imposing a maximum bond dimension. The latter constraint controls the computational cost associated with the trace of the squared MPO, keeping the calculation tractable. 

\section{Numerical data for Section III \label{AppA}}
\begin{table}[h!]
\centering
\begin{minipage}[ht]{\columnwidth}
\centering
\begin{tabular}{cc}
\toprule
$\alpha$ & number of samples \\
\midrule
1  & 51  \\ 
4  & 21 \\ 
6  & 66 \\ 
8  & 51 \\ 
10 & 21 \\ 
\bottomrule
\end{tabular}
\caption{Number of circuit realizations used for the OPEE dynamics in Fig.~\ref{fig:opee1}.}
\label{table1}
\end{minipage}
\end{table}
\newpage
\begin{table}[h!]
\begin{minipage}[t]{\columnwidth}
\centering
\begin{tabular}{cc}
\toprule
$\alpha$ & number of samples \\
\midrule
2  & 11 \\
4  & 10 \\
6  & 53 \\ 
10 & 51 \\ 
16 & 76 \\ 
30 & 24 \\ 
\bottomrule
\end{tabular}
\caption{Number of circuit realizations used for the bond dimension in Fig.~\ref{fig:bond_dimensionN100} and the OPEE dynamics in Fig.~\ref{fig:opee2}.}
\label{table2}
\end{minipage}

\vspace{0.8em}

\begin{minipage}[t]{\columnwidth}
\centering
\begin{tabular}{ccccc}
\toprule
& \multicolumn{4}{c}{$\alpha$} \\
\cmidrule(lr){2-5}
depth $(t)$ & 4 & 6 & 10 & 16 \\
\midrule
2   & 60 & 200  & 200 & 200 \\
4   & 60 & 200  & 200 & 200 \\
10  & 60 & 200  & 200 & 200 \\
20  & 60 & 200  & 200 & 200 \\
30  & 60 & 200  & 200 & 200 \\
40  & 60 & 1200 & 200 & 200 \\
55  & -- & --   & 200 & --  \\
70  & -- & 50   & 200 & 200 \\
85  & -- & --   & 100 & --  \\
100 & -- & --   & 200 & 200 \\
\bottomrule
\end{tabular}
\caption{Number of samples of OTOC data.}
\label{table3}
\end{minipage}
\end{table}
\begin{table}[h]
\centering
\begin{tabular}{cccc@{\hspace{0.8em}}ccc}
\toprule
& \multicolumn{3}{c}{$\beta_{\mathrm{OTOC}}$}
& \multicolumn{3}{c}{$\beta_{\mathrm{OPEE}}$} \\
\cmidrule(lr){2-4}\cmidrule(lr){5-7}
$\alpha$ & $t_i$ & $t_f$ & $\beta$ & $t_i$ & $t_f$ & $\beta$ \\
\midrule
4  & 10  & 40  & \num{0.59(3)}    & 8 & 36  & \num{0.525(2)}  \\
6  & 4  & 70  & \num{0.41(3)}    & 20 & 68  & \num{0.398(1)} \\
10 & 10 & 100 & \num{0.234(12)}  & 20 & 92  & \num{0.285(2)}  \\
16 & 20 & 100 & \num{0.15(4)}    & 20 & 757 & \num{0.169(1)}  \\
\bottomrule
\end{tabular}
\caption{Fitted exponents displayed in Fig. \ref{fig:otoc4} for different disorder strengths.}
\label{table4}
\end{table}
\section{$N_{sc}$ counting in OTOC computation \label{AppB}}
The gates contributing to $Z_i(t)$ are arranged in staircase circuits. For each time $t$ and pair of sites $(i,j)$, we need to evaluate the number of staircase circuits, denoted $N_{\rm sc}$. 
In the simulations of Section \ref{3.3} we have $i=N/2$ and $t\in[2\ldots N/2]$.

The gates contributing to the OTOC (highlighted in Fig. ~\ref{fig:otoc0}) are the same for pairs $(i,j)$ and $(i,j+1)$ when $j$ is odd. Therefore, it is sufficient to compute $Z_i Z_j(t)$ only for odd $j$. The operator $Z_i Z_{j+1}(t)$ is then equal to $Z_{j+1}Z_{j}Z_iZ_{j}(t)$, which simply replaces the Pauli-$Z$ operator on site $j$ with one on site $j+1$.

Then to compute the number of staircase circuits $N_{\rm{sc}}$ we consider only odd $j$. We define
\[
j_{\rm max}
\begin{cases}
i + t - 1, & j>i,\\
i - t + 1,  & j<i.
\end{cases}
\]
such that the OTOC is zero for $j > j_{\rm max}$. The number of staircase circuits is then given by the number of two-qubit gates required to connect site $i$ to site $j$:
\[
N_{\rm sc} = \frac{\left| j_{\rm max} - j\right|}{2} + 1 = 
\begin{cases}
\frac{t - \left| i-j\right| - 1}{2} + 1, & j>i,\\
\frac{t - \left| i-j\right| + 1}{2} + 1,  & j<i.
\end{cases}
\]
\section{Fidelity scaling in system size \label{AppC}}
The fidelity $F$ Eq.\eqref{fidelity} scaling in system size is readily illustrated for product unitaries. Consider 
\begin{align}
    V_{1}&=\bigotimes_{n=1}^{N}v_{n,1},
    &
    V_{2}=\bigotimes_{n=1}^{N}v_{n,2}
\end{align}
with
\begin{equation}
    v_{n,k}= \begin{pmatrix}
        \cos(\frac{x_{n,k}\pi}{2}) & \sin(\frac{x_{n,k}\pi}{2})\\
        \sin(\frac{x_{n,k}\pi}{2}) & -\cos(\frac{x_{n,k}\pi}{2})\\
    \end{pmatrix}
\end{equation}
with $n$ labeling the qubit, and $k$ the local unitary. The parameters satisfy $0\leq x_{n,k}\leq 1$.
The fidelity between $U_1$ and $U_2$ is 
\begin{equation}
    F=\dfrac{1}{2^{N}}\left| \textrm{Tr}\bigotimes_{n=1}^{N} v_{n,1}v_{n,2} \right|=\dfrac{1}{2^{N}}\left|\prod_{n=1}^{N}\textrm{Tr} \left(v_{n,1}v_{n,2}\right) \right|
\end{equation}
Assuming $x_{n,2} - x_{n,1} = \varepsilon_{n}$, the fidelity becomes
\begin{equation}
    F
    =
    \prod_{n=1}^{N}
    \cos\!\left(\frac{\varepsilon_n\pi}{2}\right).
\end{equation}

For a uniform perturbation $\varepsilon_n=\varepsilon$, this reduces to
\begin{equation}
    F
    =
    \left[
        \cos\!\left(\frac{\varepsilon\pi}{2}\right)
    \right]^N.
\end{equation}
Since
\[
    \cos\!\left(\frac{\varepsilon\pi}{2}\right)\leq 1,
\]
the fidelity decays exponentially with the system size $N$. In particular,
\begin{equation}
    F \sim F_1^N,
    \label{scalingfid}
\end{equation}
where $F_1$ denotes the single-qubit fidelity. Local errors therefore accumulate multiplicatively.
\section{Environment computation \label{AppD}}
\begin{figure*}[t]
    \includegraphics[width=\linewidth]{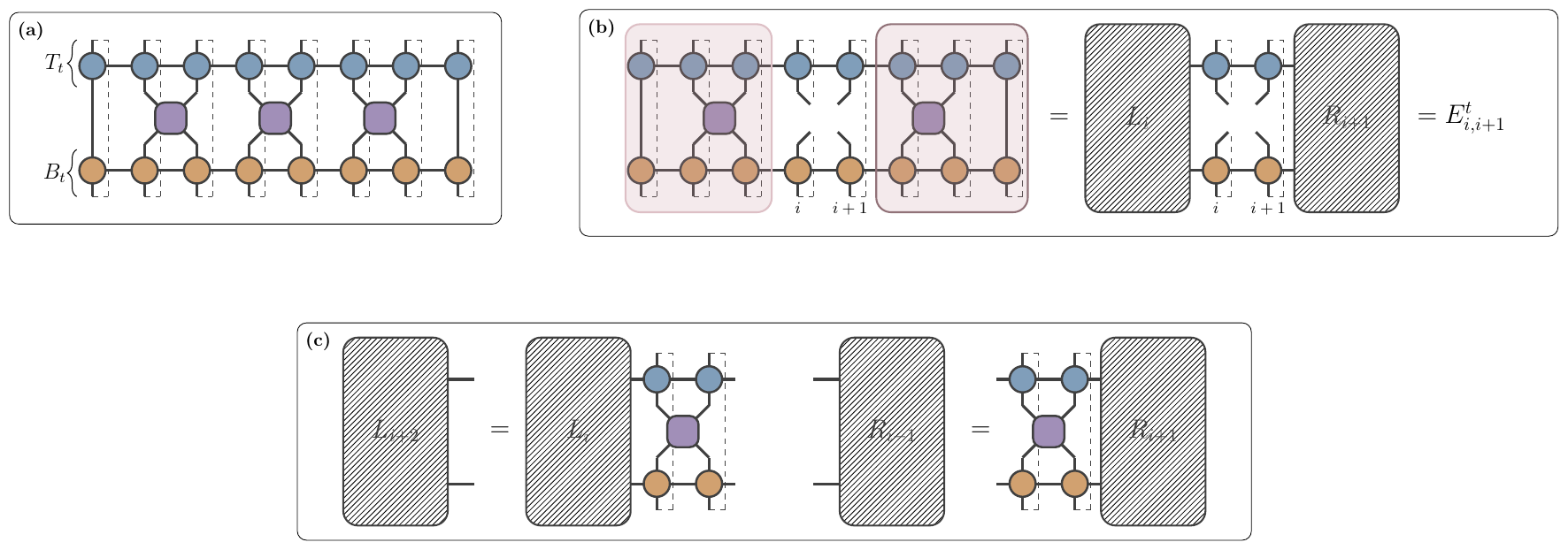}
    \caption{Tensor Network based scheme for computing the environment $E^t_{i,i+1}$. The dashed lines indicates the trace operation. $(a)$ The gates above(below) the layer $t$ are compressed into a top(bottom) MPO: $T^t$ ($B^t$). $(b)$ Contraction of the gates left of position $i$ results in the operator $L_i$, analogously, those on the right of $i+1$ are contracted into $R_{i+1}$. $(c)$ Updates of the left (right) environments in a right (left) sweep.}
    \label{fig:vcc2}
\end{figure*}
 To compute the environment $E^{t}_{i, i+1}$ we follow the scheme illustrated in Fig.\ref{fig:vcc2}. The two networks of gates located respectively above and below the layer $t$ are contracted into two MPOs $T_t$ and $B_t$, given by 
\begin{align}
   T_t&\equiv I\cdot l_{t+n}\cdot\ldots l_{t+1} \\
   B_t&\equiv l_{t-1}\cdot l_{t-2}\cdot\ldots (U^{t})^{\dagger} \ ,
\end{align}
with boundary conditions $T_{t_{\textrm{max}}}= I$, and $B_1=(U^{t})^{\dagger}$. 
The MPO representation allows control over the networks complexity. In our simulations, the truncation error is set to $10^{-7}$. 
As the sweeps in time progress, this truncation error accumulates. To mitigate this effect, the operators are reset to their boundary conditions whenever either the top or bottom boundary is reached.
The top and bottom operators are updated iteratively, by contracting a circuit layer, or its adjoint, with the current top $(T_t)$ and bottom $(B_t)$ MPOs according to:
\begin{align}
    T_{t+1}&=T_{t}\cdot l^{\dagger}_{t+1} &  B_{t+1}&=l_{t}\cdot B_{t}
\end{align}

The next step consists of computing the left $L_i$ and right $R_{i+1}$ tensors of the hole at position $(t, i, i+1)$ as shown in \figpanel{fig:vcc2}{b}
The environment $E^{t}_{i, i+1}$ results from the contraction of $L_i$, $R_{i+1}$ and the local tensors of the top $T^t$ and bottom $B^t$ MPOs acting on sites $i$ and $i+1$. 

During the inner sweep over the qubits, the left and right tensors are updated iteratively as illustrated in \figpanel{fig:vcc2}{c}. 

\section{\label{AppE} Global updates}
\begin{figure}[t]
    \centering
    \includegraphics[width=\columnwidth]{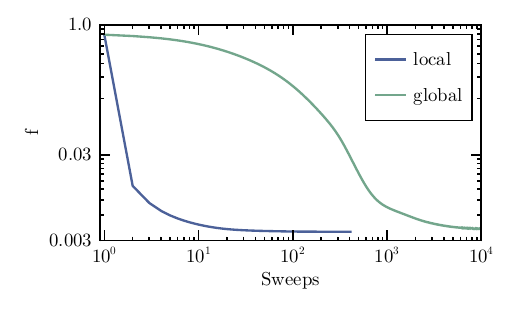}
    \caption{Comparison of local and global (RAdam) update optimization for the compression of a Floquet circuit at $\alpha=6$, and depth $t=8$ into an Ansatz of depth $D_{\rm{min}}=4$. The two solvers reach the same value of $f$, with local updates the convergence is $\approx 25$ times faster.}
    \label{fig:opts}
\end{figure}
\begin{algorithm}[t]
\caption{Riemannian adaptation of Adam solver}
\label{alg:riemannian_adam}
\SetAlgoLined
\DontPrintSemicolon
\KwIn{Initial point at iteration step $k$, $G_{k} \in \mathcal{M}$, learning rate $\alpha$, parameters $\beta_1,\beta_2$, tolerance $\varepsilon$}
\KwOut{Update $G_{k+1}$}
\BlankLine
\BlankLine
\textbf{1. Riemannian gradient computation}\;
Compute $\nabla_Rf$ at $G_{k}$ according to Eq.~\eqref{Riemgrad}\;
\BlankLine
\textbf{2. Moments updates}\;
$m_{k+1} = \beta_1\ \tilde{m}_{k}
      + (1-\beta_1)\,\nabla_R f$\; 
\BlankLine
$v_{k+1} = \beta_2\, v_{k}
      + (1-\beta_2)\,\|\nabla_R f\|^2$\;
\BlankLine
\textbf{3. Update and retraction of G}\;
$G_{k+1}=G_{k} -\alpha_{k+1} \ \dfrac{m_{k+1}}{\sqrt{v_{k+1}} + \epsilon} $
$G_{k+1} = WSV^\dagger, \qquad G_{k+1}\leftarrow WV^\dagger\in \mathcal{M}$
\BlankLine
\textbf{4. Transport $m_{k+1}$  on tangent space of current point ($\mathcal{T}_{G_{k+1}}$})\;
$\tilde{m}_{k+1} = m_{k+1} - \dfrac{1}{2}\left[ m^\dagger_{k+1} G_{k+1} - G_{k+1}^\dagger m_{k+1} \right]$\;
\end{algorithm}
Global updates consist of updating the gates in $C$ at once. Here we perform global updates of the gates using gradient descent optimization methods on unitary manifold, which have proven successful in a great number of related works \cite{10.21468/SciPostPhys.14.4.073, zhang2024scalablequantumdynamicscompilation,danna2025circuitcompression2dquantum,Kotil_2024,riemannqopt}. In our tests we use the ADAM solver \cite{riemannqopt,kingma2017adammethodstochasticoptimization}. We provide a comparison between local and global updates for our optimization problem, with the former exhibiting a faster convergence. 
Global updates are performed using the Adam optimizer adapted to the manifold of unitary matrices $\mathcal{M}$.
At each iteration, the gradient of $f$ (Eq.\eqref{logf}) w.r.t. each circuit gate is projected onto the tangent space of the unitary manifold, yielding the \emph{Riemannian gradient}~\cite{wrightbook}. The tangent space at a point $G$ is indicated as $\mathcal{T}_G(\mathcal{M})$.
The computation of the Euclidean gradient $\nabla f$ for complex-valued matrices and its projection onto the tangent space follow standard constructions; we refer the reader to Ref.~\cite{riemannqopt} for details.
$\nabla f$ with respect to one gate $G$ is proportional to the corresponding environment $E$, and reads
\begin{align}
      & \qquad\qquad\qquad\qquad \nabla f = 2\cdot\left( g\cdot h\right)^* \cdot E^*\nonumber \\ \\
   & g=-\frac{1}{2\cdot N}\frac{1}{\lvert \text{Tr}[U^{\dagger}\cdot C] \rvert^{2}} \qquad\qquad h=\text{Tr}[U^{\dagger}\cdot C]^*\nonumber
\end{align}
The Riemannian gradient $\nabla_{R}f$ is then obtained by projecting $\nabla f$ onto $\mathcal{T}_G(\mathcal{M})$,
\begin{equation}
    \nabla_{R}f=\nabla f -\frac{1}{2}\left[\nabla f^{\dagger}\cdot G - G^{\dagger}\cdot \nabla f \right] .\label{Riemgrad}
\end{equation}.
The momenta are then updated as in Algorithm~\ref{alg:riemannian_adam}. 
Compared to the standard algorithm, Adam on Riemannian manifolds needs the following modifications: 

(i) $\nabla f$ is replaced by $\nabla_R f$, (ii) The first moment $m_{k+1}$ lies in the tangent space at $G_k$; to combine it with the gradient at the next step, it is transported to the tangent space at $G_{k+1}$ before summation (step 4 of the Algorithm). (iii) The update step lies in the tangent space $\mathcal{T}_G(\mathcal{M})$ and must therefore be mapped back to the manifold via a retraction, implemented here through a singular value decomposition. 

Our implementation follows the Julia reference \textit{adam.jl} from the \textit{Optim.jl} package \cite{Optim.jl-2018}.

\section{\label{AppE1}Compression at low logarithmic fidelity density}
\begin{figure}[ht]
    \centering
    \includegraphics[width=\columnwidth]{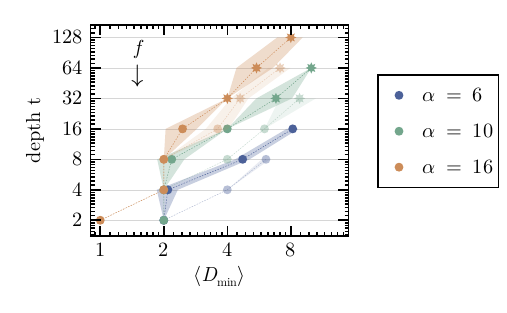}
    \caption{Exact circuit depth $(t)$ against averaged minimal depth $\langle D_{\mathrm{min}}\rangle$ achieved at stopping threshold $f=5\cdot 10^{-3}$ (darker colors) and $f=10^{-3}$ (lighter colors).}
    \label{fig:vcc_app}
\end{figure}
For completeness, in Fig.~\ref{fig:vcc_app} we show the compression data obtained at a lower stopping threshold, $f=10^{-3}$, together with the results for $f=5\cdot 10^{-3}$ studied in the main text. The figure compares the exact circuit depth $t$ with the minimal depth $\langle D_{\mathrm{min}}\rangle$ averaged over circuit realizations.

For both stopping thresholds, compression is achieved at stronger disorder, $\alpha=10$ and $\alpha=16$, showing that depth reduction in this regime is robust upon lowering the fidelity-density threshold. For $\alpha=6$, instead, the lower threshold gives $\langle D_{\mathrm{min}}\rangle \approx t$, indicating that a stricter threshold requires nearly the full circuit depth.

The less stringent threshold, $f=5\cdot 10^{-3}$, naturally allows us to obtain shallower circuits and to compress the Floquet target circuit more efficiently at longer times for every disorder strength $\alpha$ considered. The two values of $f$ correspond to two distinct global fidelities, see Eq.~(\ref{fidelity}), namely approximately $F\simeq 0.7$ for $f=5\cdot 10^{-3}$ and $F\simeq 0.9$ for $f=10^{-3}$. As pointed out in the main text, the accuracy of the optimization is assessed through physical observables, in particular through the OTOC absolute error. This test takes into account not only the error due to the logarithmic-fidelity threshold, but also the error introduced by the target approximation (Sec. \ref{4.1.2}).

\section{t-SNE Circuit visualization \label{AppF}}
\begin{figure}[ht]
    \centering
    \includegraphics[width=\columnwidth]{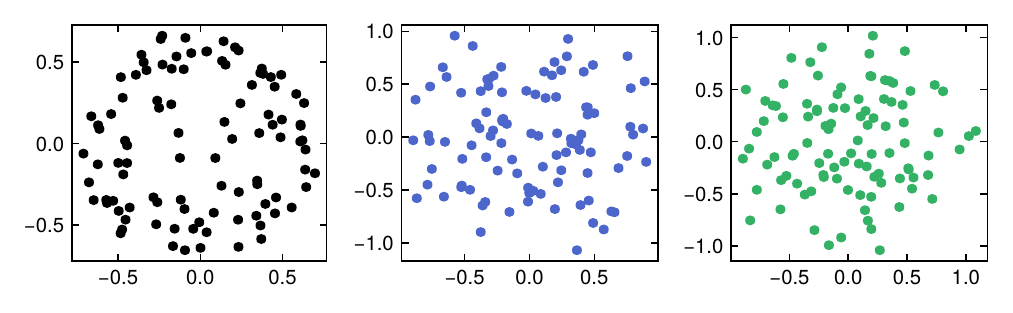}
    \includegraphics[width=\columnwidth]{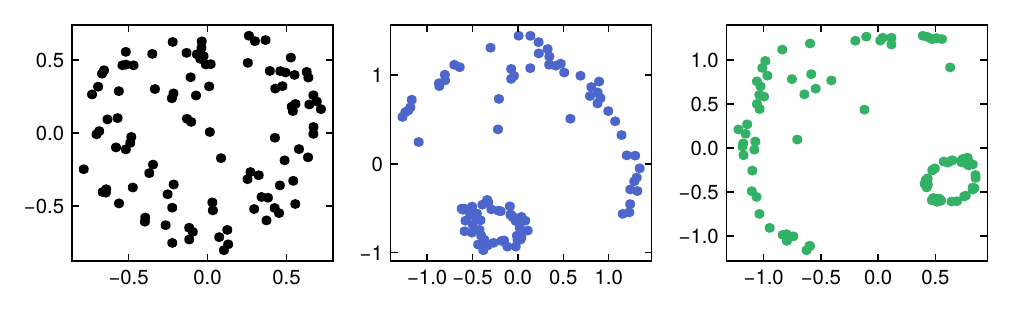}
    \caption{One period of one Floquet sample for increasing $\alpha$. From left to right $\alpha=[1,6,16]$. Plots in the first row are only for the gates in the brickwork  $(U_2)$, while those in the second row includes the layer of one-qubit gates $(U_1)$. One dot corresponds to one gate embedded in two dimension with t-SNE.}
    \label{fig:tsne}
\end{figure}
The t-SNE algorithm \cite{JMLR:v9:vandermaaten08a} provides a two-dimensional embedding of high-dimensional data. 
Figure~\ref{fig:tsne} shows the gates of one Floquet period ($U$) for $\alpha \in [1,6,16]$, with each point representing a gate: a gate composing $U$ is transformed into a real-component vector of dimension $2\cdot d^2$, where $d$ is the gate dimension. 

The first row corresponds to the brickwork layer $(U_2)$ alone, while the second includes on-site disorder (Fig.~\ref{fig:qc1}) $(U_2\cdot U_1 )$. This allows us to visualize the impact of $U_1$ to circuit complexity. Without the layer $U_1$, the gates are scattered in space, regardless of disorder strength $(\alpha)$.
Things are different when we introduce the layer $U_1$.
For weak disorder ($\alpha=1$), the distribution remains featureless -- another indicator that for this value the circuit is chaotic --. 
As $\alpha$ increases, the embedding develops structure. Gates from different layers become spatially separated, leading to a more structured embedding at strong disorder. This organization indicates a simpler optimization landscape, consistent with enhanced compressibility at strong disorder.
\bibliography{bibliography}

@article{itensor,
    title={{The ITensor Software Library for Tensor Network Calculations}},
    author={Matthew Fishman and Steven R. White and E. Miles Stoudenmire},
    journal={SciPost Phys. Codebases},
    pages={4},
    year={2022},
    publisher={SciPost},
    doi={10.21468/SciPostPhysCodeb.4},
    url={https://scipost.org/10.21468/SciPostPhysCodeb.4}
}

@article{JURECA,
author = {{J\"{u}lich Supercomputing Centre}},
title = {{JURECA: Data Centric and Booster Modules implementing the Modular Supercomputing Architecture at J\"{u}lich Supercomputing Centre}},
journal = {Journal of large-scale research facilities},
number = {A182},
volume = {7},
doi = {10.17815/jlsrf-7-182},
url = {http://dx.doi.org/10.17815/jlsrf-7-182},
year = {2021}
}

@dataset{defranco_2026_data,
  author    = {De Franco, Francesca and Kennes, Dante M. and Luitz, David J. and Rizzi, Matteo and Schmitt, Markus},
  title     = {Data for “Disorder-enhanced compression of Floquet random quantum circuits”},
  year      = {2026},
  publisher = {Zenodo},
  doi       = {10.5281/zenodo.20744506},
  url       = {https://doi.org/10.5281/zenodo.20744506}
}

@misc{DeFranco_code,
  author       = {De Franco, Francesca},
  title        = {Circuit compression code},
  year         = {2026},
  howpublished = {\url{https://github.com/frastudur/circuitcompression.git}},
  note         = {GitHub repository}
}

@article{Preskill2018quantumcomputingin,
  doi = {10.22331/q-2018-08-06-79},
  url = {https://doi.org/10.22331/q-2018-08-06-79},
  title = {Quantum {C}omputing in the {NISQ} era and beyond},
  author = {Preskill, John},
  journal = {{Quantum}},
  issn = {2521-327X},
  publisher = {{Verein zur F{\"{o}}rderung des Open Access Publizierens in den Quantenwissenschaften}},
  volume = {2},
  pages = {79},
  month = aug,
  year = {2018}
}

@article{RevModPhys.94.015004,
  title = {Noisy intermediate-scale quantum algorithms},
  author = {Bharti, Kishor and Cervera-Lierta, Alba and Kyaw, Thi Ha and Haug, Tobias and Alperin-Lea, Sumner and Anand, Abhinav and Degroote, Matthias and Heimonen, Hermanni and Kottmann, Jakob S. and Menke, Tim and Mok, Wai-Keong and Sim, Sukin and Kwek, Leong-Chuan and Aspuru-Guzik, Al\'an},
  journal = {Rev. Mod. Phys.},
  volume = {94},
  issue = {1},
  pages = {015004},
  numpages = {69},
  year = {2022},
  month = {Feb},
  publisher = {American Physical Society},
  doi = {10.1103/RevModPhys.94.015004},
  url = {https://link.aps.org/doi/10.1103/RevModPhys.94.015004}
}

@article{Suzuki1976,
  author       = {Suzuki, Masuo},
  title        = {Generalized Trotter's Formula and Systematic Approximants of Exponential Operators and Inner Derivations with Applications to Many-Body Problems},
  journal = {Communications in Mathematical Physics},
  year         = {1976},
  volume       = {51},
  number       = {2},
  pages        = {183--190},
  date         = {1976-06},
  doi          = {10.1007/BF01609348},
  issn         = {1432-0916},
  url          = {https://doi.org/10.1007/BF01609348}
}

@article{PhysRev.109.1492,
  title = {Absence of Diffusion in Certain Random Lattices},
  author = {Anderson, P. W.},
  journal = {Phys. Rev.},
  volume = {109},
  issue = {5},
  pages = {1492--1505},
  numpages = {0},
  year = {1958},
  month = {Mar},
  publisher = {American Physical Society},
  doi = {10.1103/PhysRev.109.1492},
  url = {https://link.aps.org/doi/10.1103/PhysRev.109.1492}
}

@article{PhysRevB.76.052203,
  title = {Possible experimental manifestations of the many-body localization},
  author = {Basko, D. M. and Aleiner, I. L. and Altshuler, B. L.},
  journal = {Phys. Rev. B},
  volume = {76},
  issue = {5},
  pages = {052203},
  numpages = {4},
  year = {2007},
  month = {Aug},
  publisher = {American Physical Society},
  doi = {10.1103/PhysRevB.76.052203},
  url = {https://link.aps.org/doi/10.1103/PhysRevB.76.052203}
}

@article{PhysRevB.82.174411,
  title = {Many-body localization phase transition},
  author = {Pal, Arijeet and Huse, David A.},
  journal = {Phys. Rev. B},
  volume = {82},
  issue = {17},
  pages = {174411},
  numpages = {7},
  year = {2010},
  month = {Nov},
  publisher = {American Physical Society},
  doi = {10.1103/PhysRevB.82.174411},
  url = {https://link.aps.org/doi/10.1103/PhysRevB.82.174411}
}

@ARTICLE{2016JSP...163..998I,
       author = {{Imbrie}, John Z.},
        title = "{On Many-Body Localization for Quantum Spin Chains}",
      journal = {Journal of Statistical Physics},
         year = 2016,
        month = jun,
       volume = {163},
       number = {5},
        pages = {998-1048},
          doi = {10.1007/s10955-016-1508-x},
archivePrefix = {arXiv},
       eprint = {1403.7837},
 primaryClass = {math-ph},
       adsurl = {https://ui.adsabs.harvard.edu/abs/2016JSP...163..998I}
}

@article{PhysRevB.91.081103,
  title = {Many-body localization edge in the random-field Heisenberg chain},
  author = {Luitz, David J. and Laflorencie, Nicolas and Alet, Fabien},
  journal = {Phys. Rev. B},
  volume = {91},
  issue = {8},
  pages = {081103(R)},
  numpages = {5},
  year = {2015},
  month = {Feb},
  publisher = {American Physical Society},
  doi = {10.1103/PhysRevB.91.081103},
  url = {https://link.aps.org/doi/10.1103/PhysRevB.91.081103}
}

@article{Abanin_2017,
   title={A Rigorous Theory of Many-Body Prethermalization for Periodically Driven and Closed Quantum Systems},
   volume={354},
   ISSN={1432-0916},
   url={http://dx.doi.org/10.1007/s00220-017-2930-x},
   DOI={10.1007/s00220-017-2930-x},
   number={3},
   journal={Communications in Mathematical Physics},
   publisher={Springer Science and Business Media LLC},
   author={Abanin, Dmitry and De Roeck, Wojciech and Ho, Wen Wei and Huveneers, François},
   year={2017},
   month=June, pages={809–827} }

@article{PhysRevB.105.174205,
  title = {Avalanches and many-body resonances in many-body localized systems},
  author = {Morningstar, Alan and Colmenarez, Luis and Khemani, Vedika and Luitz, David J. and Huse, David A.},
  journal = {Phys. Rev. B},
  volume = {105},
  issue = {17},
  pages = {174205},
  numpages = {20},
  year = {2022},
  month = {May},
  publisher = {American Physical Society},
  doi = {10.1103/PhysRevB.105.174205},
  url = {https://link.aps.org/doi/10.1103/PhysRevB.105.174205}
}

@article{PhysRevB.98.134204,
  title = {Localization with random time-periodic quantum circuits},
  author = {S\"underhauf, Christoph and P\'erez-Garc\'{\i}a, David and Huse, David A. and Schuch, Norbert and Cirac, J. Ignacio},
  journal = {Phys. Rev. B},
  volume = {98},
  issue = {13},
  pages = {134204},
  numpages = {16},
  year = {2018},
  month = {Oct},
  publisher = {American Physical Society},
  doi = {10.1103/PhysRevB.98.134204},
  url = {https://link.aps.org/doi/10.1103/PhysRevB.98.134204}
}

@article{v4xv-74s7,
  title = {Probing prethermal nonergodicity through measurement outcomes of monitored quantum dynamics},
  author = {Sun, Zheng-Hang and Trigueros, Fabian Ballar and Tang, Qicheng and Heyl, Markus},
  journal = {Phys. Rev. B},
  volume = {112},
  issue = {18},
  pages = {L180306},
  numpages = {7},
  year = {2025},
  month = {Nov},
  publisher = {American Physical Society},
  doi = {10.1103/v4xv-74s7},
  url = {https://link.aps.org/doi/10.1103/v4xv-74s7}
}

@article{mezzadri2006generate,
  author       = {Mezzadri, Francesco},
  title        = {How to Generate Random Matrices from the Classical Compact Groups},
  journal      = {Notices of the American Mathematical Society},
  year         = {2006},
  volume       = {54},
  number       = {5},
  pages        = {592--604},
  note         = {arXiv:math-ph/0609050},
  doi          = {10.1090/noti1279}  
}

@article{Sierant_2025,
doi = {10.1088/1361-6633/ad9756},
url = {https://doi.org/10.1088/1361-6633/ad9756},
year = {2025},
month = {jan},
publisher = {IOP Publishing},
volume = {88},
number = {2},
pages = {026502},
author = {Sierant, Piotr and Lewenstein, Maciej and Scardicchio, Antonello and Vidmar, Lev and Zakrzewski, Jakub},
title = {Many-body localization in the age of classical computing*},
journal = {Reports on Progress in Physics}
}

@article{PhysRevLett.91.147902,
  title = {Efficient Classical Simulation of Slightly Entangled Quantum Computations},
  author = {Vidal, Guifr\'e},
  journal = {Phys. Rev. Lett.},
  volume = {91},
  issue = {14},
  pages = {147902},
  numpages = {4},
  year = {2003},
  month = {Oct},
  publisher = {American Physical Society},
  doi = {10.1103/PhysRevLett.91.147902},
  url = {https://link.aps.org/doi/10.1103/PhysRevLett.91.147902}
}

@article{PhysRevB.90.174302,
  title = {Quantum quenches in the many-body localized phase},
  author = {Serbyn, Maksym and Papi\ifmmode \acute{c}\else \'{c}\fi{}, Z. and Abanin, D. A.},
  journal = {Phys. Rev. B},
  volume = {90},
  issue = {17},
  pages = {174302},
  numpages = {10},
  year = {2014},
  month = {Nov},
  publisher = {American Physical Society},
  doi = {10.1103/PhysRevB.90.174302},
  url = {https://link.aps.org/doi/10.1103/PhysRevB.90.174302}
}

@article{PhysRevA.76.032316,
  title = {Operator space entanglement entropy in a transverse Ising chain},
  author = {Prosen, Toma\ifmmode \check{z}\else \v{z}\fi{} and Pi\ifmmode \check{z}\else \v{z}\fi{}orn, Iztok},
  journal = {Phys. Rev. A},
  volume = {76},
  issue = {3},
  pages = {032316},
  numpages = {5},
  year = {2007},
  month = {Sep},
  publisher = {American Physical Society},
  doi = {10.1103/PhysRevA.76.032316},
  url = {https://link.aps.org/doi/10.1103/PhysRevA.76.032316}
}

@article{PhysRevLett.71.1291,
  title = {Average entropy of a subsystem},
  author = {Page, Don N.},
  journal = {Phys. Rev. Lett.},
  volume = {71},
  issue = {9},
  pages = {1291--1294},
  numpages = {0},
  year = {1993},
  month = {Aug},
  publisher = {American Physical Society},
  doi = {10.1103/PhysRevLett.71.1291},
  url = {https://link.aps.org/doi/10.1103/PhysRevLett.71.1291}
}

@article{PhysRevLett.72.1148,
  title = {Proof of Page's conjecture on the average entropy of a subsystem},
  author = {Foong, S. K. and Kanno, S.},
  journal = {Phys. Rev. Lett.},
  volume = {72},
  issue = {8},
  pages = {1148--1151},
  numpages = {0},
  year = {1994},
  month = {Feb},
  publisher = {American Physical Society},
  doi = {10.1103/PhysRevLett.72.1148},
  url = {https://link.aps.org/doi/10.1103/PhysRevLett.72.1148}
}

@article{Dubail_2017,
doi = {10.1088/1751-8121/aa6f38},
url = {https://doi.org/10.1088/1751-8121/aa6f38},
year = {2017},
month = {may},
publisher = {IOP Publishing},
volume = {50},
number = {23},
pages = {234001},
author = {Dubail, J},
title = {Entanglement scaling of operators: a conformal field theory approach, with a glimpse of simulability of long-time dynamics in 1+1d},
journal = {Journal of Physics A: Mathematical and Theoretical}
}

@article{PhysRevLett.111.127205,
  title = {Ballistic Spreading of Entanglement in a Diffusive Nonintegrable System},
  author = {Kim, Hyungwon and Huse, David A.},
  journal = {Phys. Rev. Lett.},
  volume = {111},
  issue = {12},
  pages = {127205},
  numpages = {5},
  year = {2013},
  month = {Sep},
  publisher = {American Physical Society},
  doi = {10.1103/PhysRevLett.111.127205},
  url = {https://link.aps.org/doi/10.1103/PhysRevLett.111.127205}
}

@article{PhysRevLett.129.170401,
  title = {Rise and Fall, and Slow Rise Again, of Operator Entanglement under Dephasing},
  author = {Wellnitz, D. and Preisser, G. and Alba, V. and Dubail, J. and Schachenmayer, J.},
  journal = {Phys. Rev. Lett.},
  volume = {129},
  issue = {17},
  pages = {170401},
  numpages = {8},
  year = {2022},
  month = {Oct},
  publisher = {American Physical Society},
  doi = {10.1103/PhysRevLett.129.170401},
  url = {https://link.aps.org/doi/10.1103/PhysRevLett.129.170401}
}

@article{SCHOLLWOCK201196,
title = {The density-matrix renormalization group in the age of matrix product states},
journal = {Annals of Physics},
volume = {326},
number = {1},
pages = {96-192},
year = {2011},
note = {January 2011 Special Issue},
issn = {0003-4916},
doi = {https://doi.org/10.1016/j.aop.2010.09.012},
url = {https://www.sciencedirect.com/science/article/pii/S0003491610001752},
author = {Ulrich Schollwöck}
}

@article{PhysRevB.95.094206,
  title = {Operator entanglement entropy of the time evolution operator in chaotic systems},
  author = {Zhou, Tianci and Luitz, David J.},
  journal = {Phys. Rev. B},
  volume = {95},
  issue = {9},
  pages = {094206},
  numpages = {15},
  year = {2017},
  month = {Mar},
  publisher = {American Physical Society},
  doi = {10.1103/PhysRevB.95.094206},
  url = {https://link.aps.org/doi/10.1103/PhysRevB.95.094206}
}

@article{PhysRevA.87.022111,
  title = {Unitary quantum gates, perfect entanglers, and unistochastic maps},
  author = {Musz, Marcin and Ku\ifmmode \acute{s}\else \'{s}\fi{}, Marek and \ifmmode \dot{Z}\else \.{Z}\fi{}yczkowski, Karol},
  journal = {Phys. Rev. A},
  volume = {87},
  issue = {2},
  pages = {022111},
  numpages = {12},
  year = {2013},
  month = {Feb},
  publisher = {American Physical Society},
  doi = {10.1103/PhysRevA.87.022111},
  url = {https://link.aps.org/doi/10.1103/PhysRevA.87.022111}
}

@article{DeChiara_2006,
doi = {10.1088/1742-5468/2006/03/P03001},
url = {https://doi.org/10.1088/1742-5468/2006/03/P03001},
year = {2006},
month = {mar},
publisher = {},
volume = {2006},
number = {03},
pages = {P03001},
author = {De Chiara, Gabriele and Montangero, Simone and Calabrese, Pasquale and Fazio, Rosario},
title = {Entanglement entropy dynamics of Heisenberg chains},
journal = {Journal of Statistical Mechanics: Theory and Experiment}
}

@article{PhysRevB.77.064426,
  title = {Many-body localization in the Heisenberg $XXZ$ magnet in a random field},
  author = {{\v Z}nidari{\v c}, Marko and Prosen, Toma{\v z} and Prelov{\v s}ek, Peter},
  journal = {Phys. Rev. B},
  volume = {77},
  issue = {6},
  pages = {064426},
  numpages = {5},
  year = {2008},
  month = {Feb},
  publisher = {American Physical Society},
  doi = {10.1103/PhysRevB.77.064426},
  url = {https://link.aps.org/doi/10.1103/PhysRevB.77.064426}
}

@article{PhysRevB.93.060201,
  title = {Extended slow dynamical regime close to the many-body localization transition},
  author = {Luitz, David J. and Laflorencie, Nicolas and Alet, Fabien},
  journal = {Phys. Rev. B},
  volume = {93},
  issue = {6},
  pages = {060201},
  numpages = {5},
  year = {2016},
  month = {Feb},
  publisher = {American Physical Society},
  doi = {10.1103/PhysRevB.93.060201},
  url = {https://link.aps.org/doi/10.1103/PhysRevB.93.060201}
}

@article{PhysRevLett.109.017202,
  title = {Unbounded Growth of Entanglement in Models of Many-Body Localization},
  author = {Bardarson, Jens H. and Pollmann, Frank and Moore, Joel E.},
  journal = {Phys. Rev. Lett.},
  volume = {109},
  issue = {1},
  pages = {017202},
  numpages = {5},
  year = {2012},
  month = {Jul},
  publisher = {American Physical Society},
  doi = {10.1103/PhysRevLett.109.017202},
  url = {https://link.aps.org/doi/10.1103/PhysRevLett.109.017202}
}

@article{PhysRevB.96.020406,
  title = {Information propagation in isolated quantum systems},
  author = {Luitz, David J. and Bar Lev, Yevgeny},
  journal = {Phys. Rev. B},
  volume = {96},
  issue = {2},
  pages = {020406},
  numpages = {5},
  year = {2017},
  month = {Jul},
  publisher = {American Physical Society},
  doi = {10.1103/PhysRevB.96.020406},
  url = {https://link.aps.org/doi/10.1103/PhysRevB.96.020406}
}

@article{LarkinOvchinnikov1969,
  author    = {A. I. Larkin and Yu. N. Ovchinnikov},
  title     = {Quasiclassical Method in the Theory of Superconductivity},
  journal   = {Sov. Phys. JETP},
  volume    = {28},
  number    = {6},
  pages     = {1200--1205},
  year      = {1969},
  note      = {Engl. transl. of Zh. Eksp. Teor. Fiz. 55, 2262--2272 (1968)},
}

@article{Lieb1972FiniteGroupVelocity,
  author  = {Lieb, Elliott H. and Robinson, Derek W.},
  title   = {The Finite Group Velocity of Quantum Spin Systems},
  journal = {Communications in Mathematical Physics},
  volume  = {28},
  number  = {3},
  pages   = {251--257},
  year    = {1972}
}

@article{MaldacenaShenkerStanford2016,
  author  = {Maldacena, Juan and Shenker, Stephen H. and Stanford, Douglas},
  title   = {A bound on chaos},
  journal = {Journal of High Energy Physics},
  year    = {2016},
  volume  = {2016},
  number  = {8},
  pages   = {106},
  doi     = {10.1007/JHEP08(2016)106},
}

@article{PhysRevX.8.031058,
  title = {Diffusive Hydrodynamics of Out-of-Time-Ordered Correlators with Charge Conservation},
  author = {Rakovszky, Tibor and Pollmann, Frank and von Keyserlingk, C. W.},
  journal = {Phys. Rev. X},
  volume = {8},
  issue = {3},
  pages = {031058},
  numpages = {28},
  year = {2018},
  month = {Sep},
  publisher = {American Physical Society},
  doi = {10.1103/PhysRevX.8.031058},
  url = {https://link.aps.org/doi/10.1103/PhysRevX.8.031058}
}

@article{PhysRevB.95.054201,
  title = {Characterizing many-body localization by out-of-time-ordered correlation},
  author = {He, Rong-Qiang and Lu, Zhong-Yi},
  journal = {Phys. Rev. B},
  volume = {95},
  issue = {5},
  pages = {054201},
  numpages = {5},
  year = {2017},
  month = {Feb},
  publisher = {American Physical Society},
  doi = {10.1103/PhysRevB.95.054201},
  url = {https://link.aps.org/doi/10.1103/PhysRevB.95.054201}
}

@article{https://doi.org/10.1002/andp.201600332,
author = {Chen, Xiao and Zhou, Tianci and Huse, David A. and Fradkin, Eduardo},
title = {Out-of-time-order correlations in many-body localized and thermal phases},
journal = {Annalen der Physik},
volume = {529},
number = {7},
pages = {1600332},
year = {2017},
doi = {https://doi.org/10.1002/andp.201600332},
url = {https://onlinelibrary.wiley.com/doi/abs/10.1002/andp.201600332}
}

@article{Bohrdt_2017,
doi = {10.1088/1367-2630/aa719b},
url = {https://doi.org/10.1088/1367-2630/aa719b},
year = {2017},
month = {jun},
publisher = {IOP Publishing},
volume = {19},
number = {6},
pages = {063001},
author = {Bohrdt, A and Mendl, C B and Endres, M and Knap, M},
title = {Scrambling and thermalization in a diffusive quantum many-body system},
journal = {New Journal of Physics}
}

@article{PhysRevX.8.021013,
  title = {Operator Hydrodynamics, OTOCs, and Entanglement Growth in Systems without Conservation Laws},
  author = {von Keyserlingk, C. W. and Rakovszky, Tibor and Pollmann, Frank and Sondhi, S. L.},
  journal = {Phys. Rev. X},
  volume = {8},
  issue = {2},
  pages = {021013},
  numpages = {19},
  year = {2018},
  month = {Apr},
  publisher = {American Physical Society},
  doi = {10.1103/PhysRevX.8.021013},
  url = {https://link.aps.org/doi/10.1103/PhysRevX.8.021013}
}

@article{PhysRevResearch.6.033062,
  title = {Scalable simulation of nonequilibrium quantum dynamics via classically optimized unitary circuits},
  author = {Causer, Luke and Jung, Felix and Mitra, Asimpunya and Pollmann, Frank and Gammon-Smith, Adam},
  journal = {Phys. Rev. Res.},
  volume = {6},
  issue = {3},
  pages = {033062},
  numpages = {19},
  year = {2024},
  month = {Jul},
  publisher = {American Physical Society},
  doi = {10.1103/PhysRevResearch.6.033062},
  url = {https://link.aps.org/doi/10.1103/PhysRevResearch.6.033062}
}

@Article{10.21468/SciPostPhys.14.4.073,
	title={Optimal compression of quantum many-body time evolution operators into brickwall circuits},
	author={Maurits S. J. Tepaske and  Dominik Hahn and David J. Luitz},
	journal={SciPost Phys.},
	volume={14},
	pages={073},
	year={2023},
	publisher={SciPost},
	doi={10.21468/SciPostPhys.14.4.073},
	url={https://scipost.org/10.21468/SciPostPhys.14.4.073}
}

@article{zhang2024scalablequantumdynamicscompilation,
  title = {Scalable quantum dynamics compilation via quantum machine learning},
  author = {Zhang, Yuxuan and Wiersema, Roeland and Carrasquilla, Juan and Cincio, Lukasz and Kim, Yong Baek},
  journal = {Phys. Rev. Res.},
  volume = {8},
  issue = {2},
  pages = {023128},
  numpages = {15},
  year = {2026},
  month = {May},
  publisher = {American Physical Society},
  doi = {10.1103/wswv-nq6d},
  url = {https://link.aps.org/doi/10.1103/wswv-nq6d}
}

@misc{danna2025circuitcompression2dquantum,
      title={Circuit compression for 2D quantum dynamics}, 
      author={Matteo D'Anna and Yuxuan Zhang and Roeland Wiersema and Manuel S. Rudolph and Juan Carrasquilla},
      year={2025},
      eprint={2507.01883},
      archivePrefix={arXiv},
      primaryClass={quant-ph},
      url={https://arxiv.org/abs/2507.01883}, 
}

@article{riemannqopt,
    author = {Le,I. N. M. and Sun,S. and Mendl,C. B.},
    title = {Riemannian quantum circuit optimization based on matrix product operators},
    journal = {Quantum 9, 1833},
    year = {2025}
}

@Article{math10060940,
AUTHOR = {Li, Wei-Ming and Ran, Shi-Ju},
TITLE = {Non-Parametric Semi-Supervised Learning in Many-Body Hilbert Space with Rescaled Logarithmic Fidelity},
JOURNAL = {Mathematics},
VOLUME = {10},
YEAR = {2022},
NUMBER = {6},
ARTICLE-NUMBER = {940},
URL = {https://www.mdpi.com/2227-7390/10/6/940},
ISSN = {2227-7390},
DOI = {10.3390/math10060940}
}

@article{Zhou_2008,
doi = {10.1088/1751-8113/41/41/412001},
url = {https://doi.org/10.1088/1751-8113/41/41/412001},
year = {2008},
month = {sep},
publisher = {},
volume = {41},
number = {41},
pages = {412001},
author = {Zhou, Huan-Qiang and Barjaktarevič, John Paul},
title = {Fidelity and quantum phase transitions},
journal = {Journal of Physics A: Mathematical and Theoretical}
}

@article{PhysRevB.79.144108,
  title = {Algorithms for entanglement renormalization},
  author = {Evenbly, G. and Vidal, G.},
  journal = {Phys. Rev. B},
  volume = {79},
  issue = {14},
  pages = {144108},
  numpages = {20},
  year = {2009},
  month = {Apr},
  publisher = {American Physical Society},
  doi = {10.1103/PhysRevB.79.144108},
  url = {https://link.aps.org/doi/10.1103/PhysRevB.79.144108}
}

@misc{kingma2017adammethodstochasticoptimization,
      title={Adam: A Method for Stochastic Optimization}, 
      author={Diederik P. Kingma and Jimmy Ba},
      year={2017},
      eprint={1412.6980},
      archivePrefix={arXiv},
      primaryClass={cs.LG},
      url={https://arxiv.org/abs/1412.6980}, 
}

@book{wrightbook,
title={Numerical Optimization},
author={Nocedal,J. and Wright,S.},
year={2006},
publisher={Springer Science and Business Media}
}

@article{Kotil_2024,
doi = {10.1088/1751-8121/ad2d6e},
url = {https://doi.org/10.1088/1751-8121/ad2d6e},
year = {2024},
month = {mar},
publisher = {IOP Publishing},
volume = {57},
number = {13},
pages = {135303},
author = {Kotil, Ayse and Banerjee, Rahul and Huang, Qunsheng and Mendl, Christian B},
title = {Riemannian quantum circuit optimization for Hamiltonian simulation},
journal = {Journal of Physics A: Mathematical and Theoretical}
}

@article{Optim.jl-2018,
  author  = {Mogensen, Patrick Kofod and Riseth, Asbj{\o}rn Nilsen},
  title   = {Optim: A mathematical optimization package for {Julia}},
  journal = {Journal of Open Source Software},
  year    = {2018},
  volume  = {3},
  number  = {24},
  pages   = {615},
  doi     = {10.21105/joss.00615}
}

@article{Gibbs2025deepcircuit,
  doi = {10.22331/q-2025-07-09-1789},
  url = {https://doi.org/10.22331/q-2025-07-09-1789},
  title = {Deep {C}ircuit {C}ompression for {Q}uantum {D}ynamics via {T}ensor {N}etworks},
  author = {Gibbs, Joe and Cincio, Lukasz},
  journal = {{Quantum}},
  issn = {2521-327X},
  publisher = {{Verein zur F{\"{o}}rderung des Open Access Publizierens in den Quantenwissenschaften}},
  volume = {9},
  pages = {1789},
  month = jul,
  year = {2025}
}

@article{JMLR:v9:vandermaaten08a,
  author  = {Laurens van der Maaten and Geoffrey Hinton},
  title   = {Visualizing Data using t-SNE},
  journal = {Journal of Machine Learning Research},
  year    = {2008},
  volume  = {9},
  number  = {86},
  pages   = {2579--2605},
  url     = {http://jmlr.org/papers/v9/vandermaaten08a.html}
}

@article{google2025quantum,
  author  = {{Google Quantum AI and Collaborators}},
  title   = {Observation of constructive interference at the edge of quantum ergodicity},
  journal = {Nature},
  volume  = {646},
  pages   = {825--830},
  year    = {2025},
  doi     = {10.1038/s41586-025-09526-6},
  url     = {https://doi.org/10.1038/s41586-025-09526-6}
}

@article{Hosur2016,
	title = {Chaos in quantum channels},
	volume = {2016},
	issn = {1029-8479},
	url = {https://doi.org/10.1007/JHEP02(2016)004},
	doi = {10.1007/JHEP02(2016)004},
	number = {2},
	urldate = {2026-04-14},
	journal = {Journal of High Energy Physics},
	author = {Hosur, Pavan and Qi, Xiao-Liang and Roberts, Daniel A. and Yoshida, Beni},
    month={Feb},
	year = {2016},
	pages = {4}
}

@article{Schnaack2019,
  title = {Tripartite information, scrambling, and the role of Hilbert space partitioning in quantum lattice models},
  author = {Schnaack, Oskar and B\"olter, Niklas and Paeckel, Sebastian and Manmana, Salvatore R. and Kehrein, Stefan and Schmitt, Markus},
  journal = {Phys. Rev. B},
  volume = {100},
  issue = {22},
  pages = {224302},
  numpages = {8},
  year = {2019},
  month = {Dec},
  publisher = {American Physical Society},
  doi = {10.1103/PhysRevB.100.224302},
  url = {https://link.aps.org/doi/10.1103/PhysRevB.100.224302}
}

@article{Boelter2022,
  title = {Scrambling and many-body localization in the XXZ chain},
  author = {B\"olter, Niklas and Kehrein, Stefan},
  journal = {Phys. Rev. B},
  volume = {105},
  issue = {10},
  pages = {104202},
  numpages = {11},
  year = {2022},
  month = {Mar},
  publisher = {American Physical Society},
  doi = {10.1103/PhysRevB.105.104202},
  url = {https://link.aps.org/doi/10.1103/PhysRevB.105.104202}
}

@article{Mi_2021,
   title={Information scrambling in quantum circuits},
   volume={374},
   ISSN={1095-9203},
   url={http://dx.doi.org/10.1126/science.abg5029},
   DOI={10.1126/science.abg5029},
   number={6574},
   journal={Science},
   publisher={American Association for the Advancement of Science (AAAS)},
   author={Mi, Xiao and Roushan, Pedram and Quintana, Chris and Mandrà, Salvatore and Marshall, Jeffrey and Neill, Charles and Arute, Frank and Arya, Kunal and Atalaya, Juan and Babbush, Ryan and Bardin, Joseph C. and Barends, Rami and Basso, Joao and Bengtsson, Andreas and Boixo, Sergio and Bourassa, Alexandre and Broughton, Michael and Buckley, Bob B. and Buell, David A. and Burkett, Brian and Bushnell, Nicholas and Chen, Zijun and Chiaro, Benjamin and Collins, Roberto and Courtney, William and Demura, Sean and Derk, Alan R. and Dunsworth, Andrew and Eppens, Daniel and Erickson, Catherine and Farhi, Edward and Fowler, Austin G. and Foxen, Brooks and Gidney, Craig and Giustina, Marissa and Gross, Jonathan A. and Harrigan, Matthew P. and Harrington, Sean D. and Hilton, Jeremy and Ho, Alan and Hong, Sabrina and Huang, Trent and Huggins, William J. and Ioffe, L. B. and Isakov, Sergei V. and Jeffrey, Evan and Jiang, Zhang and Jones, Cody and Kafri, Dvir and Kelly, Julian and Kim, Seon and Kitaev, Alexei and Klimov, Paul V. and Korotkov, Alexander N. and Kostritsa, Fedor and Landhuis, David and Laptev, Pavel and Lucero, Erik and Martin, Orion and McClean, Jarrod R. and McCourt, Trevor and McEwen, Matt and Megrant, Anthony and Miao, Kevin C. and Mohseni, Masoud and Montazeri, Shirin and Mruczkiewicz, Wojciech and Mutus, Josh and Naaman, Ofer and Neeley, Matthew and Newman, Michael and Niu, Murphy Yuezhen and O’Brien, Thomas E. and Opremcak, Alex and Ostby, Eric and Pato, Balint and Petukhov, Andre and Redd, Nicholas and Rubin, Nicholas C. and Sank, Daniel and Satzinger, Kevin J. and Shvarts, Vladimir and Strain, Doug and Szalay, Marco and Trevithick, Matthew D. and Villalonga, Benjamin and White, Theodore and Yao, Z. Jamie and Yeh, Ping and Zalcman, Adam and Neven, Hartmut and Aleiner, Igor and Kechedzhi, Kostyantyn and Smelyanskiy, Vadim and Chen, Yu},
   year={2021},
   month=dec, pages={1479–1483} 
   }

@article{PhysRevB.95.155129,
  title = {Stability and instability towards delocalization in many-body localization systems},
  author = {De Roeck, Wojciech and Huveneers, Fran\ifmmode \mbox{\c{c}}\else \c{c}\fi{}ois},
  journal = {Phys. Rev. B},
  volume = {95},
  issue = {15},
  pages = {155129},
  numpages = {14},
  year = {2017},
  month = {Apr},
  publisher = {American Physical Society},
  doi = {10.1103/PhysRevB.95.155129},
  url = {https://link.aps.org/doi/10.1103/PhysRevB.95.155129}
}

@article{PhysRevB.108.134204,
  title = {Internal clock of many-body delocalization},
  author = {Evers, Ferdinand and Modak, Ishita and Bera, Soumya},
  journal = {Phys. Rev. B},
  volume = {108},
  issue = {13},
  pages = {134204},
  numpages = {15},
  year = {2023},
  month = {Oct},
  publisher = {American Physical Society},
  doi = {10.1103/PhysRevB.108.134204},
  url = {https://link.aps.org/doi/10.1103/PhysRevB.108.134204}
}
\end{document}